\documentclass[trackchanges,modern]{aastex7}

\usepackage{bm}
\usepackage{amsmath}
\usepackage{graphicx}
\usepackage[noend]{algpseudocode}
\usepackage{tikz}
\usetikzlibrary{shapes.geometric, arrows}
\usetikzlibrary{positioning}

\tikzstyle{startstop} = [rectangle, rounded corners, 
minimum width=2.0cm, 
minimum height=0.75cm,
text centered, 
draw=black, 
text width=2.0cm,
fill=gray!30]

\tikzstyle{io} = [trapezium, 
trapezium stretches=true, 
trapezium left angle=70, 
trapezium right angle=110, 
minimum width=2.3cm, 
minimum height=1cm, 
text centered, 
text width=2.3cm, 
draw=black, fill=#1]

\tikzstyle{int} = [trapezium, 
trapezium stretches=true, 
trapezium left angle=70, 
trapezium right angle=110, 
minimum width=3cm, 
minimum height=1cm, 
text centered, 
text width=3cm, 
draw=black, fill=red!30]

\tikzstyle{process} = [rectangle, 
minimum width=3cm, 
minimum height=1cm, 
text centered, 
text width=3cm, 
draw=black, 
fill=green!30]

\tikzstyle{decision} = [diamond, 
minimum width=3cm, 
minimum height=1cm, 
text centered, 
draw=black, 
fill=yellow!30]
\tikzstyle{arrow} = [thick,->,>=stealth]

\newcounter{algorithm}
\renewcommand{\thealgorithm}{\arabic{algorithm}}

\newenvironment{aalgorithm}[1][]%
{%
  \refstepcounter{algorithm}%
  \par\smallskip
  \noindent\hrule height 0.4pt
  \vspace{0.3ex}
  \noindent\textbf{Algorithm~\thealgorithm.}%
  \if\relax\detokenize{#1}\relax\else\ \textbf{#1}\fi
  \par\vspace{0.2ex}
  \noindent\hrule height 0.4pt
  \vspace{0.4ex}
}%
{%
  \vspace{0.4ex}
  \noindent\hrule height 0.4pt
  \par\smallskip
}

\newcommand{\tn}[1]{\textnormal{#1}}

\DeclareMathOperator*{\argmax}{arg\,max}

\begin{document}

\title{Discovering pulsars in compact binaries with a hidden Markov model}

\author[orcid=0000-0002-6547-2039]{Joseph O'Leary}
\affiliation{School of Physics, University of Melbourne, Parkville, VIC 3010, Australia.}
\affiliation{Australian Research Council Centre of Excellence for Gravitational Wave Discovery (OzGrav), Parkville, VIC 3010, Australia.}
\email[show]{joe.oleary@unimelb.edu.au}  

\author[orcid=0000-0002-1769-6097]{Liam Dunn} 
\affiliation{School of Physics, University of Melbourne, Parkville, VIC 3010, Australia.}
\affiliation{Australian Research Council Centre of Excellence for Gravitational Wave Discovery (OzGrav), Parkville, VIC 3010, Australia.}
\email{liamd@student.unimelb.edu.au}

\author[orcid=0000-0003-4642-141X]{Andrew Melatos}
\affiliation{School of Physics, University of Melbourne, Parkville, VIC 3010, Australia.}
\affiliation{Australian Research Council Centre of Excellence for Gravitational Wave Discovery (OzGrav), Parkville, VIC 3010, Australia.}
\email{amelatos@unimelb.edu.au}

\begin{abstract}
Discovering radio pulsars in compact binaries, whose orbital periods $P_{\rm b}$ satisfy $P_{\rm b} \lesssim 1 \, \rm{day}$, is computationally challenging, because the time-dependent pulse frequency $f_{\rm p}(t)$ is strongly Doppler modulated by the binary motion. Here we present a new, fast, semi-coherent detection scheme based on a hidden Markov model (HMM)  combined with a maximum likelihood matched filter, the Schuster periodogram. The HMM scheme complements traditional acceleration searches by dividing $f_{\rm p}(t)$ into piecewise-constant blocks and tracking the block-to-block evolution efficiently using dynamic programming. Monte Carlo simulations show that the new method can detect compact binaries with flux densities $S \geq 0.50 \, \rm{mJy}$ and orbital periods $P_{\rm b} \geq 0.012 \, \rm{day}$ under observing conditions (e.g.\ cadence) typical of radio pulsar surveys, with and without impulsive, narrowband radio frequency interference. The new method is fast; it employs the classic Viterbi algorithm to solve the HMM recursively. The central processing unit run time scales nominally as $T_{\rm run} \approx 2.8 \, N_B (N_T/10^2) (N_Q \ln N_Q/10^4 \ln 10^4) \, {\rm s}$ for $N_B$ subbands, $N_T$ coherent segments, and $N_Q$ frequency bins. 
\end{abstract}


\section{Introduction}

Pulsar surveys \citep{Manchester_2001,Morris_2002,Kramer_2003,Faulkner_2004,Hobbs_2004,Manchester_2005} have led to the discovery of approximately $3380$ pulsars, of which $439$ are in binaries, and $155$ are in compact binaries whose orbital periods $P_{\rm{b}}$ satisfy $0.05 \lesssim P_{\rm{b}}/(1 \, \rm{day}) \lesssim 1$; see the Australian Telescope National Facility (ATNF) pulsar catalogue for an up-to-date compilation of population statistics.\footnote{\href{https://www.atnf.csiro.au/research/pulsar/psrcat/}{https://www.atnf.csiro.au/research/pulsar/psrcat/} \label{FN:ATNFCat}} Recent surveys \citep{Nan_2011,Bailes_2020,Han_2021,Ridolfi_2021,Chen_2023,Wang_2023} combine new data from latest-generation radio telescopes with archival data reprocessed using upgraded hardware and software, e.g.\ graphics processing units \citep{Morello_2019,Crawford_2021,Sengar_2023,wongphechauxsorn_2023}, distributed volunteer computing projects such as Einstein@Home \citep{Knispel_2013, Lazarus_2016}, novel search algorithms such as image pattern recognition \citep{Zhu_2014}, and techniques from artificial intelligence such as neural networks \citep{Eatough_2010}. 

Standard search pipelines in blind pulsar surveys usually apply Fourier techniques to dedispersed and barycentred time series generated using a range of trial dispersion measures. The pipelines search the resulting Fourier spectra for significant features, i.e.\ coherent pulsations; see Chapters 5 and 6 in \cite{Lorimer_2005} for overviews on instrumentation and detection techniques, respectively. Compact binaries with $0.05 \lesssim P_{\rm{b}}/(1 \, \rm{day}) \lesssim 1$  present particularly acute computational and observational challenges, because the pulse frequency $f_{\rm{p}}(t)$ is strongly Doppler modulated by the binary motion, spreading the signal across an extended comb of Doppler sidebands in the Fourier power spectrum, and reducing the signal-to-noise ratio per sideband \citep{Johnston_1991}. 

One popular technique in binary pulsar searches is time-domain resampling \citep{Camilo_2000,Eatough_2013}, where the dedispersed time series is transformed to the pulsar's rest frame using the standard Doppler formula; see Equation (6.16) in \cite{Lorimer_2005}. It is possible in principle to remove completely the signal modulation due to binary motion, provided the pulsar radial velocity along the line of sight $V_1(t)$ is known a priori, a challenge in blind pulsar surveys. When the orbital parameters are unknown, a common strategy to mitigate the loss of sensitivity is to assume a constant line-of-sight orbital acceleration, i.e.\ $\tn{d}V_1/\tn{d}t = a_1$, and trial a range of $a_1$ values with each dedispersed time series; see Section 6.2.1 in \cite{Lorimer_2005} for a discussion on selecting trial orbital acceleration values, and Section 2 in \cite{Camilo_2000} for a guide on implementing the technique in practice. Care must be taken when selecting $a_1$ values, so as to avoid computational overhead and maximise search sensitivity. Although acceleration searches \citep{Johnston_1991} are powerful binary detection techniques, they are restricted to scenarios, where the total observation time $T_{\rm{obs}}$ accounts for a small fraction of the orbital period, e.g. $T_{\rm{obs}}\lesssim 0.1 P_{\rm{b}}$ \citep{Ransom_2003}. Hence the pulsar must be bright, as the minimum flux density required to detect a pulsar scales $\propto T^{-1/2}_{\rm{obs}}$ \citep{Camilo_2000}. 

In addition to constant line-of-sight acceleration searches, several other techniques have been developed to address the challenges associated with pulsar detection, binary or otherwise. (i) Constant ``jerk'' searches \citep{Anderson_2018}, with  $\tn{d}a_1/\tn{d}t = j_1$, are the logical extension of the aforementioned acceleration search technique, increasing both search sensitivity (by allowing longer integrations e.g.\ $T_{\rm{obs}} \lesssim 0.15 P_{\rm{b}}$) and computation time, e.g.\ by a factor of $\sim 80$ in the case of pulsar PSR J1748--2446am \citep{Anderson_2018}. (ii) Fully coherent techniques based on matched filters search over three (circular) or five (elliptical) Keplerian orbital elements \citep{Balakrishnan_2022}, leveraging distributed volunteer computing projects, e.g.\ Einstein@Home. Fully coherent searches discovered previously undetected pulsars in archival Parkes Multibeam Pulsar Survey data \citep{Manchester_2001,Knispel_2013} and Pulsar Arecibo L-band Feed Array survey data \citep{Cordes_2006,Allen_2013}; see Table 7 in \cite{Balakrishnan_2022} for runtime comparisons when searching over three and five Keplerian orbital elements  on simulated data. (iii) Semi-coherent sideband searches \citep{Ransom_2003} complement acceleration searches, targeting ``ultracompact'' binaries, with $P_{\rm{b}} \lesssim 0.16 \, {\rm{day}}$. We refer the reader to Chapter 6 in \cite{Lorimer_2005} for overviews of the various pulsar search techniques in the time and frequency domains. 

Hidden Markov model (HMM) algorithms \citep{Baum_1966,Rabiner_1989,Streit_1990} offer a powerful, statistical framework for frequency tracking in low signal-to-noise conditions, where the frequency wanders either stochastically or deterministically, e.g.\ due to binary motion. The reader is referred to Chapter 7 in \cite{Quinn_2001} for an overview of frequency tracking using a HMM. The basic idea is to relate the discrete transitions of an unobservable (``hidden'') state -- the unknown frequency $f_{\rm p}(t)$ of a yet-to-be-discovered pulsar, for example -- to a set of timing observations via a detection statistic such as the Fourier transform \citep{Suvorova_2016}. The state transitions are modeled probabilistically as a Markov chain. In the astrophysical context, HMMs have been employed across numerous applications including continuous gravitational-wave searches \citep{Suvorova_2016,Suvorova_2017,Abbott_2017,Sun_2019,Abbott_2019,Middleton_2020,Melatos_2021,Beniwal_2021} and pulsar glitch detection \citep{Melatos_2020,Dunn_2022a,Dunn_2023}, complementing numerous, practical applications in electrical engineering \citep{Rabiner_1989,Xie_1991,Krishnamurthy_2001}.  Several of the foregoing publications target sources in binary systems, where the HMM harvests the signal power in every orbital Doppler sidebands \citep{Suvorova_2016,Suvorova_2017,Abbott_2017,Abbott_2019,Middleton_2020,Melatos_2021,Abbott_2022,Abbott_2022b,Vargas_2023a,Vargas_2023b}.

In this paper, we demonstrate how to combine a HMM with time-series data from a typical radio pulsar survey to efficiently discover compact binary pulsars with $0.05 \lesssim P_{\rm b} / (1 \, {\rm day}) \lesssim 1$. The method is validated deliberately with synthetic data in order to quantify its performance (e.g.\ the minimum flux density for a detection) systematically under controlled conditions. It will be applied to real data in a forthcoming paper. We elect to model the pulse frequency $f_{\rm{p}}(t)$ using a HMM for two reasons: (i) HMMs are computationally efficient, e.g.\ \cite{Suvorova_2016} demonstrated a runtime improvement of up to three orders of magnitude over other, semi-coherent algorithms for continuous gravitational wave tracking from the low-mass X-ray binary, Scorpius X$-$1  [see Table V in \cite{Suvorova_2016} for further details]; and (ii) HMMs have been  applied successfully to frequency estimation problems, when samples are abundant but the signal-to-noise ratio is low -- the situation relevant to searches for compact binary pulsars.

We emphasize that the approach adopted herein does not supersede well-established methods for binary pulsar detection, e.g.\ time-domain resampling \citep{Camilo_2000,Eatough_2013} or fully coherent techniques \citep{Balakrishnan_2022}. HMM-based methods are a new, semi-coherent pulsar detection tool to be deployed in tandem with existing techniques and software, e.g.\ \textsc{presto} \citep{Ranson_2011}. They are especially suited to blind searches for compact binaries by increasing the efficiency of trialling many constant acceleration values \citep{Johnston_1991} or Keplerian orbital elements \citep{Allen_2013,Knispel_2013,Balakrishnan_2022} by leveraging a dynamic programming algorithm. 

The paper is structured as follows. In Section \ref{Sec:HMMAlg} we summarise the components of a HMM in the context of discovering binary pulsars. The method is validated on simulated radio survey data for a single, representative test source in Section \ref{Sec:Validation}. In Section \ref{Sec:PerformanceTests}, we quantify how the binary orbital elements affect the accuracy of the HMM, as well as place limits on the minimum flux density required for a detection. In Section \ref{Sec:CompCost}, we show that the HMM is relatively efficient computationally and present empirical scalings for the run time as a function of key search parameters. Astrophysical implications are canvassed briefly in Section \ref{Sec:Conclusions}, together with a note on generalizing the HMM to real data, the topic of a future paper.

\section{Pulse frequency tracking with a HMM} \label{Sec:HMMAlg}
A HMM solved recursively with the classic Viterbi algorithm \citep{Viterbi_1967,Rabiner_1989} provides an efficient, statistical framework to infer the maximum likelihood evolution of a hidden state variable, related probabilistically to a time-ordered sequence of observations via a detection statistic. In the binary pulsar context, the observations are a dedispersed and barycentred radio intensity time series, the hidden state is the pulse frequency $f_{\rm{p}}(t)$, and the detection statistic is the Fourier power, discussed in detail below. In Section \ref{SubSec:HMMForm} we review briefly the components of an arbitrary HMM. In Sections \ref{SubSec:Inputs}--\ref{SubSec:PriorTrans} we specialize the components, introduced in Section \ref{SubSec:HMMForm}, to the specific task of binary pulsar detection. An overview of the binary pulsar detection workflow is given in Section \ref{SubSec:Workflow}. We explain how to set the detection threshold in Section \ref{SubSec:Threshold}. The Viterbi algorithm logic and pseudocode are summarized in Appendix \ref{App:Viterbi} for the convenience of the reader.

\subsection{Finite-state automaton} \label{SubSec:HMMForm}

We implement a HMM as a probabilistic finite-state automaton.  At time $t \in \{t_0, \hdots , t_{N_T} \}$, the automaton occupies the hidden state $q(t) \in \{q_1, \hdots, q_{N_Q} \}$. Similarly, we assume that the system is observable, with observation $o(t) \in \{o_1, \hdots, o_{N_O} \}$.\footnote{The observation $o(t)$ at time $t$ need not be discrete. Generalizations to continuous observation densities are discussed in Section IV of \cite{Rabiner_1989} as well as \cite{Liporace_1982}, \cite{Juang_1985}, and \cite{Juang_1986}.} Here we assume the automaton is Markovian, i.e.\ the hidden state transition probability from time $t_n$ to $t_{n+1}$ depends only on the hidden state $q(t_n)$ at time $t_n$.

Given the hidden state and observation sequences, denoted respectively by $Q=\{q(t_0),\hdots,q(t_{N_T})\}$ and $O = \{o(t_0),\hdots o(t_{N_T}) \}$, the most likely path $Q^{*} = \{ q^{*}(t_0),\hdots,q^{*}(t_{N_T}) \}$ maximizes $P(Q|O)$, viz.
\begin{align} \label{Eq:MaxPathQ}
    Q^{*}(O) =& \,  \argmax_{Q} \big[ L_{o(t_{N_T})q(t_{N_T})}A_{q(t_{N_T})q(t_{N_T - 1})} \times \hdots \notag \\ 
    & \times L_{o(t_{1})q(t_{1})}A_{q(t_{1})q(t_{0})} \Pi_{q(t_0)} \big].
\end{align}
The HMM components $M = \{A, L, \Pi \}$ in Equation (\ref{Eq:MaxPathQ}) are defined as follows:
\begin{equation} \label{Eq:TransProb}
    A_{q_jq_i} = \textnormal{Pr}[q(t_{n+1}) = q_j| q(t_{n}) = q_i], 
\end{equation}
\begin{equation} \label{Eq:EmissProb}
    L_{o_jq_i} = \textnormal{Pr}[o(t_n) = o_j| q(t_n) = q_i],
\end{equation}
\begin{equation} \label{Eq:InitProb}
    \Pi_{q_i} = {\rm{Pr}}[q(t_0) = q_i].
\end{equation}
The $N_Q \times N_Q$ and $N_O \times N_Q$ matrices $A_{q_j q_i}$ and $L_{o_j q_i}$ are defined in terms of conditional probabilities and are called the transition and emission probability matrices, respectively. The prior vector $\Pi_{q_i}$ is the probability that the system occupies the hidden state $q_i$ at time $t_0$. In practice we work with logarithms to avoid numerical issues, so the product on the right-hand side of Equation (\ref{Eq:MaxPathQ}) becomes a sum. 

Ultimately, the choice of variables in Equations (\ref{Eq:TransProb})--(\ref{Eq:InitProb}) is specific to the problem and lies with the analyst. The choice appropriate for this paper is defined and justified in Section \ref{SubSec:Inputs}. By way of additional background, the reader is referred to Section 3 of \cite{Melatos_2020} and Section 2 of \cite{Melatos_2021} for  how to formulate related but different problems in pulsar astronomy in terms of a HMM, namely pulsar glitch detection and continuous gravitational-wave searches. 

\subsection{Pulsar survey data and HMM mapping} \label{SubSec:Inputs}
A typical pulsar survey generates a dedispersed and barycentred radio intensity time series $x(t')$ per sky pointing, of duration $T_{\rm{obs}}$ and comprising $N$ samples, with $t' \in \{t'_0, \cdots, t'_{N-1} \}$ and $t'_{N-1} = t'_0+T_{\rm obs}$. The time series is divided into $m = 0, \hdots, N_T$ coherent segments $o(t_m)$, each of duration $T_{\rm{coh}} = T_{\rm{obs}}/N_T$ (with $t_m \leq t' \leq t_m + T_{\rm coh}$) and comprising $N_{\rm coh} = f_{\rm samp} T_{\rm coh}$ samples, where $f_{\rm samp}$ equals the sampling frequency of the recording system, which collects the radio intensity data $x(t')$. That is, the $m$-th coherent segment yields an $N_{\rm coh}$-dimensional measurement vector of the form $o(t_m) = [ x(t'_{m'}=t_m), \dots, x(t'_{m'+N_{\rm coh}-1}) ]$, where $m'$ labels the radio intensity sample coincident with $t_m$. Specifically, $t \in \{t_0, \hdots , t_{N_T} \}$ labels the time coordinate adopted internally by the HMM, and $t' \in \{t'_0, \hdots , t'_{N-1} \}$ labels the times at which the radio intensity data are collected, with $t_r \neq t'_s$ for some (and maybe all) $r$ and $s$ in general. 

In this paper we track the hidden pulse frequency $q(t) = f_{\rm{p}}(t)$ of a yet-to-be discovered pulsar, as observed in the Solar System barycentre. We discretize $q(t)$ into $N_Q$ frequency bins, whose widths $\propto T_{\rm coh}^{-1}$ are set judiciously to avoid $f_{\rm p}(t)$ wandering secularly (due to spin down and binary motion) or stochastically (due to timing noise) by more than one bin from $t_n$ to $t_{n+1}$. The discretization recipe is discussed in Section \ref{SubSec:DataProducts} as well as by \cite{Suvorova_2016}.

\subsection{Frequency domain intermediate data products and emission probability matrix} \label{SubSec:DataProducts}
Traditional binary pulsar searches construct a detection statistic from the discrete Fourier transform of the radio intensity time series $x(t'_n)$ \citep{Burns_1969,Hankins_1975,Bhattacharya_1998,Lorimer_2005,Lyne_2012}. We adopt the same approach here when constructing the detection statistic and hence the emission probability matrix in a coherent segment. By working in the frequency (Fourier) domain, we are led to make certain choices when discretizing the hidden states, e.g.\ when choosing the frequency bin width. The discretization strategy, and the associated emission probability matrix, are described in this section.

The hidden states (frequency bins) are contained within a search band $B_{\rm min} \leq f \leq B_{\rm max}$, whose bounds $B_{\rm min}$ and $B_{\rm max}$ are specified by the analyst at their discretion based on prior astronomical expectations. The search band is divided into $N_B$ subbands of width $\Delta f_B = (B_{\rm max} - B_{\rm min})/N_B$. The $j$-th subband spans $f_{0,j} \leq f \leq f_{0,j}+\Delta f_B$, with $1\leq j \leq N_B$. The number of subbands $N_B$ adopted in the validation test in Section \ref{Sec:Validation} as well as the performance tests in Section \ref{Sec:PerformanceTests} is discussed in Section \ref{SubSec:FrequencyTracking} and reported in the bottom section of Table \ref{tab:SourceParameters}. 

The segments of radio intensity data $o(t_m)$ defined in Section \ref{SubSec:Inputs} are heterodyned, placing the middle frequency of each subband at zero Hz, and downsampled to the Nyquist sampling rate $f_{\rm{Nyq}} < f_{\rm{samp}}$, reducing the number of samples per segment to $\hat{N}_{\rm{coh}} = N_{\rm{coh}}f_{\rm{Nyq}}/f_{\rm{samp}}$. That is, the $m$-th heterodyned,  downsampled, coherent segment is a $N_{\rm coh}$-dimensional data vector of the form $\hat{o}(t_m) = [\hat{x}(\hat{t}_{\hat{m}}=t_m), \dots, \hat{x}(\hat{t}_{\hat{m}+\hat{N}_{\rm coh}-1})]$ with  $\hat{x}(\hat{t}_{\hat{m}}) = \exp[- 2 \pi i (f_{0,j} + \Delta f_B/2) \hat{t}_{\hat{m}}] $, where $\hat{m}$ labels the radio intensity sample coincident with $t_{m}$, and $\hat{t} \in \{\hat{t}_0, \cdots, \hat{t}_{\hat{N}_{\rm{coh}}-1}\}$ denotes the downsampled time coordinate.

The emission probability matrix is constructed from the discrete Fourier transform $\hat{X}_{m,k}$ of the heterodyned and downsampled radio intensity data, defined as
\begin{equation}\label{Eq:FFT}
    \hat{X}_{m, k} =  \sum^{\hat{N}_{\rm{coh}}-1}_{l=0}\hat{x}_{m,l}\exp{(-2 \pi i k l/\hat{N}_{\rm{coh}})}.
\end{equation}
In Equation (\ref{Eq:FFT}),  $\hat{X}_{m,k}$ and $\hat{x}_{m,l}$ are introduced for brevity and denote respectively the $k$-th Fourier component ($1 \leq k \leq \hat{N}_{\rm{coh}}/2$) and $l$-th heterodyned and downsampled radio intensity sample $\hat{x}_{m,l} = \hat{x}_{m}(\hat{t}_l)$ associated with the $m$-th coherent data segment. 

The HMM searches for the most probable track in the time-frequency plane, visualized as a spectrogram composed of $N_Q$ frequency bins multiplied by $N_T$ discrete time bins. The probability that the HMM observes $o(t_m)$ while in hidden state $q(t_m)$ is encoded in the components of the emission probability matrix, viz. 
\begin{equation}\label{Eq:EmissionSpec}
    L_{o(t_m)q_k} \propto \exp (P_{m,k}),
\end{equation}
where
\begin{equation}\label{Eq:Power}
    P_{m,k} = \left|\hat{X}_{m,k}\right|^2/(\hat{\sigma}_{m}^2 \hat{N}_{\rm{coh}})
\end{equation}
is the normalized Schuster periodogram \citep{Bretthorst_1988,Bayley_2019}, and is equivalent to the log-likelihood that the signal frequency lies at the center of the $k$-th frequency bin [$f_{p,k} - \Delta f_{B}/2, f_{p,k} + \Delta f_{B}/2$]. In Equation (\ref{Eq:Power}), $\hat{\sigma}^2_{m}$ denotes the variance of the $m$-th heterodyned and downsampled coherent data segment. The number of discrete time bins $N_T$ is a key input into the HMM. In the binary pulsar context, $N_T$ is controlled by the observation length $T_{\rm obs}$ as well as the binary parameters. A recipe for selecting $N_T$ in the validation and performance tests in Sections \ref{Sec:Validation} and \ref{Sec:PerformanceTests} is given in Appendix \ref{App:TimeBins}.

The periodogram is the maximum likelihood (maximized over the unknown amplitude) matched filter for a sinusoidal signal without taking into account the amplitude and frequency modulation from the Earth's diurnal rotation and annual revolution \citep{jaranowski_1998}, which the photon time-of-arrival barycentering procedure addresses \citep{Edwards_2006}. Equation (\ref{Eq:Power}) is not unique. The analyst is entitled to replace Equation (\ref{Eq:Power}) with other forms of $L_{o(t_m) q_k}$, based on a detection statistic other than a maximum likelihood matched filter, if they prefer.

\subsection{Selecting $T_{\rm coh}$}\label{SubSec:Tcoh}

The length of the coherent timescale $T_{\rm coh}$ is a key input into the HMM. An important feature of frequency tracking with a HMM is the connection between the frequency drift $\Delta f_{\rm drift}$ induced by the binary motion, the Fourier frequency bin width $\Delta f_{\rm W}$, and the coherent timescale $T_{\rm coh}$. Specifically, $\Delta f_{\rm W} \propto T_{\rm coh}^{-1}$ is set by $T_{\rm coh}$, the latter being judiciously picked to satisfy
    \begin{equation}\label{Eq:fdrift}
        \Delta f_{\rm{drift}} = \int^{t + T_{\rm coh}}_{t} d \Tilde{t} \, |d f_{\rm p}/d \Tilde{t}|
    \end{equation}
    and 
    \begin{equation}
        \Delta f_{\rm{drift}} \leq a \Delta f_{\rm W}. \label{Eq:DriftBins}
    \end{equation}
    In Equation (\ref{Eq:DriftBins}), $a$ is the maximum number of frequency bins that $f_{\rm p}(t)$ drifts between coherent segments, i.e. $\ D = \Delta f_{\rm drift}/\Delta f_{\rm W} \leq a = 1$ in the present application. Assuming that $d f_{\rm p}/d \Tilde{t}$ is approximately constant between $t$ and $t+T_{\rm coh}$, one has \citep{Chandler_2003,Lorimer_2005}
    \begin{equation}\label{Eq:TCoh}
        T_{\rm coh} \leq \left(\frac{2 \pi f_{\rm p} \beta}{P_{\rm b}}\right)^{-1/2},
    \end{equation}
    where we introduce the dimensionless parameter $\beta = V_1 \sin i /c$, where the orbital velocity, orbital inclination, and speed of light in vacuum are denoted by $V_1$, $i$, and $c$, respectively. An equivalent, alternative approach to selecting $N_T = T_{\rm obs}/T_{\rm coh}$ is presented in Appendix \ref{App:TimeBins} for the convenience of the reader.

\subsection{Signal model and transition probability matrix}\label{SubSec:PriorTrans}

We model the pulse frequency $f_{\rm{p}}(t)$ as an unbiased random walk in which $f_{\rm{p}}(t)$ transitions between discrete states, e.g.\ from $q(t_n)$ to $q(t_{n+1})$, with probabilities defined according to $A_{q_j q_i}$ in Equation (\ref{Eq:TransProb}). The transition probabilities are given by
\begin{equation}\label{Eq:TransSpec}
  A_{q_{i - 1} \, q_i} = A_{q_i q_i} = A_{q_{i + 1} \, q_i} = 1/3,  
\end{equation}
with the remaining entries being zero, i.e.\ we assume $f_{\rm{p}}(t)$ transitions by $-1$, $0$, or $+1$ frequency bins with equal probability at each discrete time step $t_n$. We emphasize that the signal model above is not derived uniquely from first principles and is merely an approximation to the actual (unknown) evolution of $f_{\rm{p}}(t)$ for a real pulsar \citep{Suvorova_2016}.

The magnetic dipole braking, i.e.\ secular spin down, of rotation-powered pulsars \citep{Goldreich_1969,Ostriker_1969} is linearly superposed with at least two additional spin wandering contributions, namely stochastic timing noise and deterministic Doppler shift; see Chapter 8 of  \cite{Lorimer_2005} for a brief summary of intrinsic pulsar timing noise. Timing noise manifests as a quasi-random walk in the rms residuals of $\phi_{\rm p}(t)$, $f_{\rm p}(t)$, or $df_{\rm p}/dt$ \citep{Boynton_1972,Groth_1975,Groth_1975b,Groth_1975c,Helfand_1980}. It has a red Fourier power spectrum $S(f) \sim f^{-\gamma}$ (with $2 \lesssim \gamma \lesssim 6$), implying a process (whose astrophysical origin is unknown) autocorrelated on timescales of hours to years \citep{Deeter_1982,Boynton_1984,Deeter_1984,Cordes_1985,DAlessandro_1997,Hobbs_2006b,Hobbs_2010,Melatos_2014,Parthasarathy_2019,Lower_2020}.

Numerous pulsar timing and gravitational wave studies have quantified timing noise using stability parameters, e.g.\ Equation (2) of \cite{Arzoumanian_1994}; Allan-variance-like statistics, e.g.\ Equation (11) of \cite{Matsakis_1997}; phenomenological scalings, e.g.\ Equation (6) of \cite{Shannon_2010}; and power spectral density methods, e.g.\ Equation (26) of \cite{Beniwal_2021}, among others \citep{Lasky_2015}. As just one example, consider the timing noise root-mean-square (rms) amplitude $2 \lesssim \sigma_{\rm{TN,meas}}(T = 10 \, \rm{yr})/(100 \, \rm{ns}) \lesssim 6$ measured for five pulsars over a 10 yr observation span, the details of which are given in Table 3 of \cite{Shannon_2010}. The foregoing measurements correspond to a timing-noise-induced frequency drift $\Delta f_{{\rm{TN}},T} \sim f_{\rm p} \, \sigma_{\rm TN, meas}/T \sim 10^{-14} \, \rm{Hz}$ for $T = 10 \, \rm{yr}$. Accordingly, one expects the timing-noise-induced frequency drift per coherent segment, $\Delta f_{{\rm{TN}},T_{\rm coh}}$, to satisfy $\Delta f_{\rm drift} \gg \Delta f_{{\rm{TN}},T} \gg \Delta f_{{\rm{TN}},T_{\rm coh}}$, where $\Delta f_{\rm drift}$  typically satisfies $10^{-5} \lesssim \Delta f_{\rm drift}/(1 \, \rm{Hz}) \lesssim 10^{-3}$. Specifically, in the interval $|t_{n+1}-t_n| \sim 10^3 \, \rm{s}$ between coherent segments, stochastic timing noise is expected to be negligible compared to the orbital Doppler shift for most pulsars, and $f_{\rm p}(t)$ evolves secularly to a good approximation from $f_{\rm p}(t_n)$ to $f_{\rm p}(t_{n+1})$ due to the Doppler shift, in a manner consistent with Equation (\ref{Eq:TransSpec}). The stochastic transition matrix in Equation (\ref{Eq:TransSpec}) is flexible enough to handle the resulting uncertainty in the secular orbital motion as well as any genuinely stochastic timing noise which is present, if the frequency bin width is set judiciously according to the recipe in Section \ref{SubSec:Tcoh}. This obviates the need to explicitly search over orbital parameters, in contrast with non-HMM analyses \citep{Allen_2013,Knispel_2013,Balakrishnan_2022}.

The signal model, Equation (\ref{Eq:TransSpec}), is widely adopted across numerous applications including continuous gravitational wave searches \citep{Suvorova_2016,Suvorova_2017,Abbott_2017,Sun_2019,Abbott_2019,Middleton_2020,Melatos_2021,Beniwal_2021} and pulsar glitch detection \citep{Melatos_2020,Dunn_2022a,Dunn_2023}. In the context of discovering pulsars in compact binaries, Equation (\ref{Eq:TransSpec}) is validated through controlled injections in the present manuscript for the first time, the details of which are given in Section \ref{Sec:Validation}. In the context of continuous gravitational wave searches and pulsar glitch detection, Equation (\ref{Eq:TransSpec}) has been validated through controlled and blind injections into synthetic Gaussian as well as real, non-Gaussian LIGO data; see (for example) Figure 1 of \cite{Suvorova_2016}, Figure 3 of \cite{Suvorova_2017}, Figures 2 and 3 of \cite{Sun_2018}, and Figure F1 of \cite{Melatos_2020}. Algorithms based on Equation (\ref{Eq:TransSpec}) detected all the blind injections in the Scorpius X$-$1 Mock Data Challenge; see \cite{Messenger_2015} and Sections 5 of \cite{Suvorova_2016} and \cite{Suvorova_2017} respectively. Most recently, \cite{Carlin_2025} validated Equation (\ref{Eq:TransSpec}) for various classes of stochastic signal models and compared its performance with other semi-coherent algorithms, e.g.\ based on cross-correlation.

In practice, we do not know a priori what frequency bin the pulsar signal occupies at time $t = t_0$. Hence we assign equal probabilities to all $N_Q$ frequency bins, viz.
\begin{equation}\label{Eq:PriorSpec}
    \Pi_{q(t_0)} = N_{Q}^{-1},
\end{equation}
over the search band $B_{\rm min} \leq q(t_0) \leq B_{\rm max}$ defined in Section \ref{SubSec:FrequencyTracking}.

Experience across numerous electrical engineering and astronomy applications teaches, that $P(Q|O)$ is insensitive to (i) the form of $A_{q_jq_i}$, provided the dynamics of $f_{\rm{p}}(t)$ are captured approximately between successive timesteps $t_{n} \leq t \leq t_{n+1}$ \citep{Quinn_2001,Suvorova_2016,Melatos_2020}; and (ii) adopting a uniform prior, because $\Pi_{q(t_0)}$ accounts for only one out of $2(N_T + 1) \gg 1$ multiplicative factors of what is generally a large product, defined by the right-hand side of Equation (\ref{Eq:MaxPathQ}) \citep{Suvorova_2016,Suvorova_2017,Abbott_2017,Melatos_2020}. Accordingly, we persevere with Equations (\ref{Eq:TransSpec}) and (\ref{Eq:PriorSpec})  as a first pass at the problem, partly driven by their successful use in other, similar astronomical applications \citep{Sun_2018,Bayley_2019,Sun_2019,Melatos_2020,Vargas_2023a,Vargas_2023b}, and partly for the sake of simplicity.

\subsection{HMM workflow} \label{SubSec:Workflow}
We employ the Viterbi algorithm to recursively solve the HMM and determine $Q^*$, the hidden state sequence that maximizes $P(Q|O)$, given $O$. Viterbi paths for which $\mathcal{L} = \ln P(Q^*|O)$ exceeds an analyst-specified detection threshold, discussed in Section \ref{SubSec:Threshold} below, are regarded as potential pulsar discoveries, which merit further analysis. 

\begin{figure}
\centering
\begin{tikzpicture}[node distance=2cm]

\node (start) [startstop] {Start};
\node (pro1) [process, below of=start] {Compute $P_{m,k}$ for each segment};
\node (in1) [io=red!30, right of=pro1, xshift=2cm] {
$N_T$ segments of DFTs, $\hat{X}_{m,k}$};
\node (in2) [io=blue!10, below of=pro1] {$N_Q \times N_T$ spectrogram};
\node (pro2) [process, below of=in2] {Run Viterbi algorithm};
\node (in3) [io=blue!10, below of=pro2] {$\mathcal{L} = \ln P(Q^*|O)$};
\node (dec1) [decision, below of=in3, yshift=-0.75cm] {$\mathcal{L} > \mathcal{L}_{\rm{th}}$?};

\node (stop1) [startstop, below of=dec1, xshift=-2.0cm] {Reject $Q^*$};
\node (stop2) [startstop, below of=dec1, xshift= 2.0cm] {Potential detection};

\draw [arrow] (start) -- (pro1);
\draw [arrow] (in1) -- (pro1);
\draw [arrow] (pro1) -- (in2);
\draw [arrow] (in2) -- (pro2);
\draw [arrow] (pro2) -- (in3);
\draw [arrow] (in3) -- (dec1);
\draw [arrow] (dec1) -- node[anchor=south west, xshift=-0.45cm] {No} (stop1);
\draw [arrow] (dec1) -- node[anchor=south east, xshift= 0.45cm] {Yes} (stop2);

\end{tikzpicture}
\caption{Workflow of the binary pulsar search pipeline for a single subband. The start and end points of the pipeline are in gray ovals. Processes are reported in green rectangles. Inputs and outputs are reported as red and blue parallelograms, respectively. Decision points are drawn as yellow diamonds. The acronym DFT stands for discrete Fourier transforms. For the tests in Section \ref{Sec:Validation}, the workflow is repeated for $N_B = 100$ subbands from 50 Hz to 1050 Hz.} 
\label{Fig:FlowChart}
\end{figure}
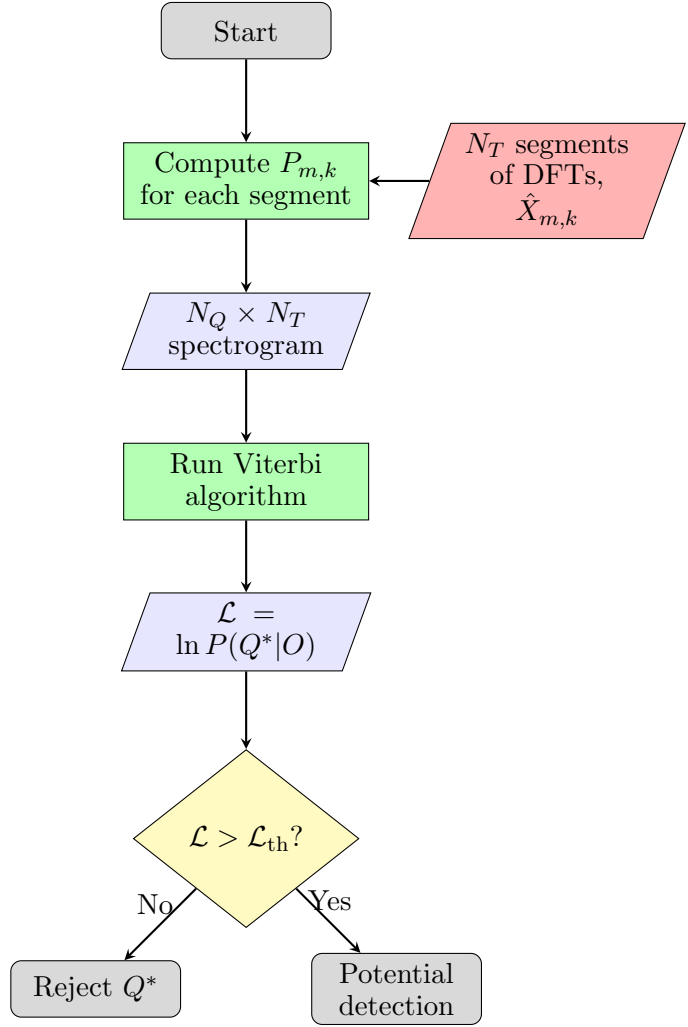

The workflow of the pulsar search over a single subband is summarized in Figure \ref{Fig:FlowChart}. Variations of Figure \ref{Fig:FlowChart} [e.g.\  Figures 1 and  2 in \cite{Abbott_2019} and \cite{Abbott_2022b}, respectively] and Algorithm \ref{Alg:Viterbi} [e.g.\ in Section II.D in \cite{Suvorova_2016} as well as appendices A in \cite{Melatos_2020} and \cite{Melatos_2021}] appear in other related applications. They are summarized here for the convenience of the reader to assist with reproducibility and because it is the first time an HMM solved by the Viterbi algorithm has been applied to binary pulsar searches. The Viterbi algorithm logic and pseudocode are summarized in Appendix \ref{App:Viterbi}.

\subsection{Detection threshold}\label{SubSec:Threshold}
It remains to define a suitable detection threshold for identifying pulsar candidates. The reader is referred to Appendix A in \cite{Abbott_2022} for a detailed discussion about detection thresholds as well as comparisons between two commonly employed strategies for threshold selection, namely the exponential tail method (employed here) and the percentile method. 

Consider the $j$-th subband with $1 \leq j \leq N_{B}$. The Viterbi algorithm in Section \ref{SubSec:Workflow} returns $\mathcal{L} = P(Q^*|O)$ for the $N_Q$ paths terminating somewhere in the $j$-th subband. This raises an important question: what likelihood threshold $\mathcal{L}_{\rm{th}}$ should $\mathcal{L}$ exceed to be considered a possible signal? The answer is not unique. It depends on the false alarm probability $\alpha'$ per subband, that the analyst is prepared to tolerate. As a starting point, we calibrate ${\cal L}_{\rm th}$ assuming Gaussian measurement noise. A systematic study of the impact of non-Gaussian noise artifacts, e.g.\ radio frequency interference from terrestrial sources, is postponed to future work; some preliminary tests are conducted in Appendix \ref{App:HMMRFI}. We refer the reader to Section \ref{SubSec:SyntheticData} for complete details about generating noisy synthetic pulsar survey data using the \texttt{tempo2}, \texttt{presto}, and \texttt{simulatesearch} software packages \citep{Edwards_2006,Hobbs_2006,Ranson_2011,Luo_2022}. 

Let us compute the noise-only probability density function (PDF) $p({\cal L})$ of ${\cal L}$ in the absence of a pulsar signal. Recall that given a time series of pure Gaussian noise denoted by $n(t)$, its real and imaginary discrete Fourier components, calculated according to Equation (\ref{Eq:FFT}), are also normally distributed. Similarly, the associated Fourier power coefficients $P_{m,k}$ (normalized by $\hat{\sigma}_{m}^2 \hat{N}_{\rm coh}$), calculated according to Equation (\ref{Eq:Power}), follow an exponential PDF, with $p(P_{m,k}) = \exp(-P_{m,k})$ \citep{Lorimer_2005}. Accordingly, we construct $p(\mathcal{L})$ by analyzing $N_{\rm real}$ Monte-Carlo realizations of an $N_{T}$ times $N_Q$ spectrogram with the Viterbi algorithm in Appendix \ref{App:Viterbi}. We retain all $N_Q$ times $N_{\rm real}$ log-likelihood estimates $\mathcal{L}$ to construct $p(\mathcal{L})$.\footnote{Including $\mathcal{L} = \ln P(Q|O)$ for $Q \neq Q^*$, i.e.\ nonmaximal paths, when constructing $p(\mathcal{L})$ does not change its shape or alter $\mathcal{L}_{\rm th}$ appreciably \citep{Abbott_2022}.}  The elements of the noise-only time-frequency spectrogram are populated using the \texttt{numpy} function \texttt{random.exponential}.\footnote{\href{https://numpy.org/doc/2.1/reference/random/generated/numpy.random.exponential.html}{https://numpy.org/doc/2.1/reference/random/}} 

We observe empirically that $p(\mathcal{L}>\mathcal{L}_{\rm tail})$ has an exponentially distributed tail in noise,
\begin{equation}\label{Eq:TailDist}
    p(\mathcal{L} > \mathcal{L}_{\rm tail}) = k \lambda \exp [-\lambda (\mathcal{L} - \mathcal{L_{\rm{tail}}})],
\end{equation}
where $k = N_{\rm{tail}}/(N_{\rm{real}}N_{Q})$ is the fraction of samples used to construct $p(\mathcal{L} > \mathcal{L}_{\rm tail})$, $N_{\rm{tail}}$ is the number of samples in $p(\mathcal{L} > \mathcal{L}_{\rm{tail}})$, $\mathcal{L}_{\rm{tail}}$ is a likelihood cut-off which must be determined empirically, and we substitute the maximum likelihood estimate
\begin{equation}\label{Eq:MLELambda}
    \Tilde{\lambda} = \frac{N_{\rm{tail}}}{\sum^{N_{\rm{tail}}}_{i=1} (\mathcal{L}_i - \mathcal{L}_{\rm{tail}})},
\end{equation}
for $\lambda$ in Equation (\ref{Eq:TailDist}) assuming that the $\mathcal{L}$ samples are independent. The probability that $\mathcal{L}$ exceeds $\mathcal{L}_{\rm{th}}$ due to random fluctuations is given by
\begin{equation}\label{Eq:ProbSingleL}
    \alpha = \int^{\infty}_{\mathcal{L}_{\rm{th}}} d \mathcal{L} \, p(\mathcal{L}>\mathcal{L}_{\rm tail}), 
\end{equation}
where $\alpha$ is related to the probability of false alarm $\alpha'$ per subband, viz. 
\begin{equation}\label{eq:Probrelation}
    \alpha' = 1 - (1 - \alpha)^{N_{Q}}.
\end{equation}
We estimate the likelihood threshold,
\begin{equation}\label{Eq:L_th}
    \mathcal{L}_{\rm{th}} = \mathcal{L}_{\rm{tail}} -\Tilde{\lambda}^{-1}\log \{ N_{\rm{real}}N_{Q}  [1 - (1 - \alpha')^{1/N_{Q}}]/N_{\rm{tail}}\},
\end{equation}
by inverting Equation (\ref{eq:Probrelation}) for $\alpha$, and combining with Equation (\ref{Eq:ProbSingleL}). The uncertainty $\sigma_{\mathcal{L}_{\rm th}}$ associated with Equation (\ref{Eq:L_th}) is derived analytically using the first-order delta method in Appendix \ref{App:sigma} and is given by the square root of Equation (\ref{Eq:LthVariance}). Equation (\ref{Eq:L_th}) coincides with Equation (A4) in \cite{Abbott_2022} and Equation (13) in \cite{Knee_2023}.

\section{Validation with synthetic data} \label{Sec:Validation}

In this section, we orient the reader through a binary pulsar detection validation test conducted on synthetic data. The test serves as a worked example, which illustrates how the detection scheme in Section \ref{Sec:HMMAlg} operates in practice, and prefigures the fuller suite of systematic performance tests in Section \ref{Sec:PerformanceTests}. The binary signal and orbital parameters of a representative test source are laid out in Section \ref{SubSec:TestSource}. In Section \ref{SubSec:SyntheticData}, we present a step-by-step guide on how to create the synthetic data by injecting a binary pulsar signal with frequency $f_{\rm{p}}(t)$ into Gaussian radiometer data $n(t)$ and generating a dedispersed and barycentred radio intensity time series using the \texttt{tempo2},\footnote{\href{https://bitbucket.org/psrsoft/tempo2}{https://bitbucket.org/psrsoft/tempo2} \label{FN:tempo2}} \texttt{presto},\footnote{\href{https://github.com/scottransom/presto}{https://github.com/scottransom/presto} \label{FN:presto}} and \texttt{simulatesearch}\footnote{\href{https://bitbucket.csiro.au/projects/psrsoft/repos/simulatesearch}{https://bitbucket.csiro.au/projects/psrsoft/repos/simulatesearch}\label{FN:SS}} software packages \citep{Edwards_2006,Hobbs_2006,Ranson_2011,Luo_2022}.  In Section \ref{SubSec:Idealizations} we discuss the approximations made in creating the synthetic data according to the recipe in Section \ref{SubSec:SyntheticData}. The performance of the HMM in the absence of a putative pulsar signal, i.e.\ noise-only characterization, is quantified in Section \ref{SubSec:NoiseOnly}. A worked example of compact binary pulsar detection with a HMM, using the data generated in Section \ref{SubSec:SyntheticData}, is presented in Section \ref{SubSec:FrequencyTracking}. 

\subsection{Representative test source} \label{SubSec:TestSource}

As a representative test source we consider a binary pulsar whose signal, rotational, and orbital parameters emulate those of PSR J1953$+$1844, the millisecond pulsar with the shortest known orbital period, with $P_{\rm b}=0.037 \, {\rm days}$ \citep{Nan_2011,Jiang_2019,Pan_2023}. The source parameters are reported in the top section of Table \ref{tab:SourceParameters} and include the average flux density $S$ (units: mJy), fractional pulse width at 50\% peak flux density $W_{50}$ (units: dimensionless), pulse frequency $f_{\rm{p,inj}}$ (units: Hz), orbital period $P_{\rm{b}}$ (units: days), projected semi-major axis $a \sin i/c$ (units: lt-s), and orbital eccentricity $e_{\rm{b}}$ (units: dimensionless) where $i$ denotes the orbital inclination. We refer the reader to the ATNF pulsar catalogue (see Footnote \ref{FN:ATNFCat} for details) as well as to the \texttt{python} ATNF query interface \texttt{psrqpy} \citep{Pitkin_2018} for summaries of additional source parameters, e.g.\ right ascension and declination of PSR J1953+1844, not reported in Table \ref{tab:SourceParameters}.  

\subsection{Generating synthetic pulsar survey data} \label{SubSec:SyntheticData}

High-time resolution radio survey data are affected by several measurement noise processes. Examples include thermal electron and sky background fluctuations, flicker and jitter noise, and radio frequency interference (RFI) \citep{Press_1978,Lee_2012,Wang_2015,Lentati_2016,Luo_2022}. In this paper, we focus on additive Gaussian noise, generated synthetically using the \texttt{simulateSystemNoise} subroutine of the \texttt{simulatesearch} software package \citep{Luo_2022}. Specifically, we generate zero-mean radiometer noise $n(t) \sim \mathcal{N}(0, \sigma^2)$, whose rms amplitude $\sigma$ is described by the canonical radiometer equation \citep{Lorimer_2005}
\begin{equation}
    \sigma = T_{\rm{sys}}G_{\rm{sys}}^{-1} (n_{\rm{p}} \, \Delta f_{\rm{sys}} \, t_{\rm{samp}})^{-1/2}.
\end{equation}
The system temperature $T_{\rm{sys}}$ (units: K), telescope gain $G_{\rm{sys}}$ (units: $\rm{K \, Jy^{-1}}$), receiver bandwidth $\Delta f_{\rm{sys}}$ (units: MHz), sampling time $t_{\rm{samp}}$, and number of polarizations $n_{\rm{p}}$ are reported in the middle section of Table \ref{tab:SourceParameters}. The foregoing parameters are adopted to emulate a Parkes ``Murriyang'' multibeam system survey with 1-bit sampling and 96 frequency channels, operating at a central frequency of 1374 MHz, details of which can be found in \cite{Edwards_2001}, \cite{Manchester_2001}, and \cite{Rane_2016}.

We employ the \texttt{tempo2} software package \citep{Edwards_2006,Hobbs_2006} to approximate the time-resolved output of a complete pulsar timing model. That is, we generate a \texttt{tempo2}-style predictor to approximate the pulse phase $\phi_{\rm{p}}(t)$ and pulse frequency $f_{\rm{p}}(t)$ of the representative test source in Table \ref{tab:SourceParameters}. We refer the reader to Section 7.2 and Equations (28)--(32) of \cite{Hobbs_2006} for details of \texttt{tempo2}'s predictive mode as well as a step-by-step guide on approximating the time-resolved output of pulsar timing models using two-dimensional Chebyshev basis functions and polynomials; see also Sections 2 and 3 of \cite{Edwards_2006} for overviews of pulsar timing models and their associated accuracy estimates. The \texttt{tempo2}-generated predictive polynomial is passed to \texttt{simulatesearch}'s \texttt{simulateComplexPsr} subroutine, converting the approximate pulsar timing solution output by \texttt{tempo2}'s \texttt{pred} function into a format compatible with the noise-only data discussed in the previous paragraph. We inject the simulated pulsar signal into the noise-only radiometer data using \texttt{simulatesearch}'s \texttt{createSearchFile} subroutine, the output of which is a PSRFITS search mode data file \citep{Hotan_2004}. The final data product analyzed by the HMM in Section \ref{Sec:HMMAlg} is a dedispersed and barycentred radio intensity time series $x(t')$, generated using \texttt{presto}'s \texttt{prepdata} function. Once $x(t')$ is generated, it is ingested by the HMM and processed according to the steps outlined in Section \ref{SubSec:DataProducts}. We refer the reader to Footnotes \ref{FN:tempo2}, \ref{FN:presto}, and \ref{FN:SS} for details about downloading and installing the foregoing software packages. 

\begin{figure}
\centering{
\includegraphics[width=0.75\columnwidth, keepaspectratio]{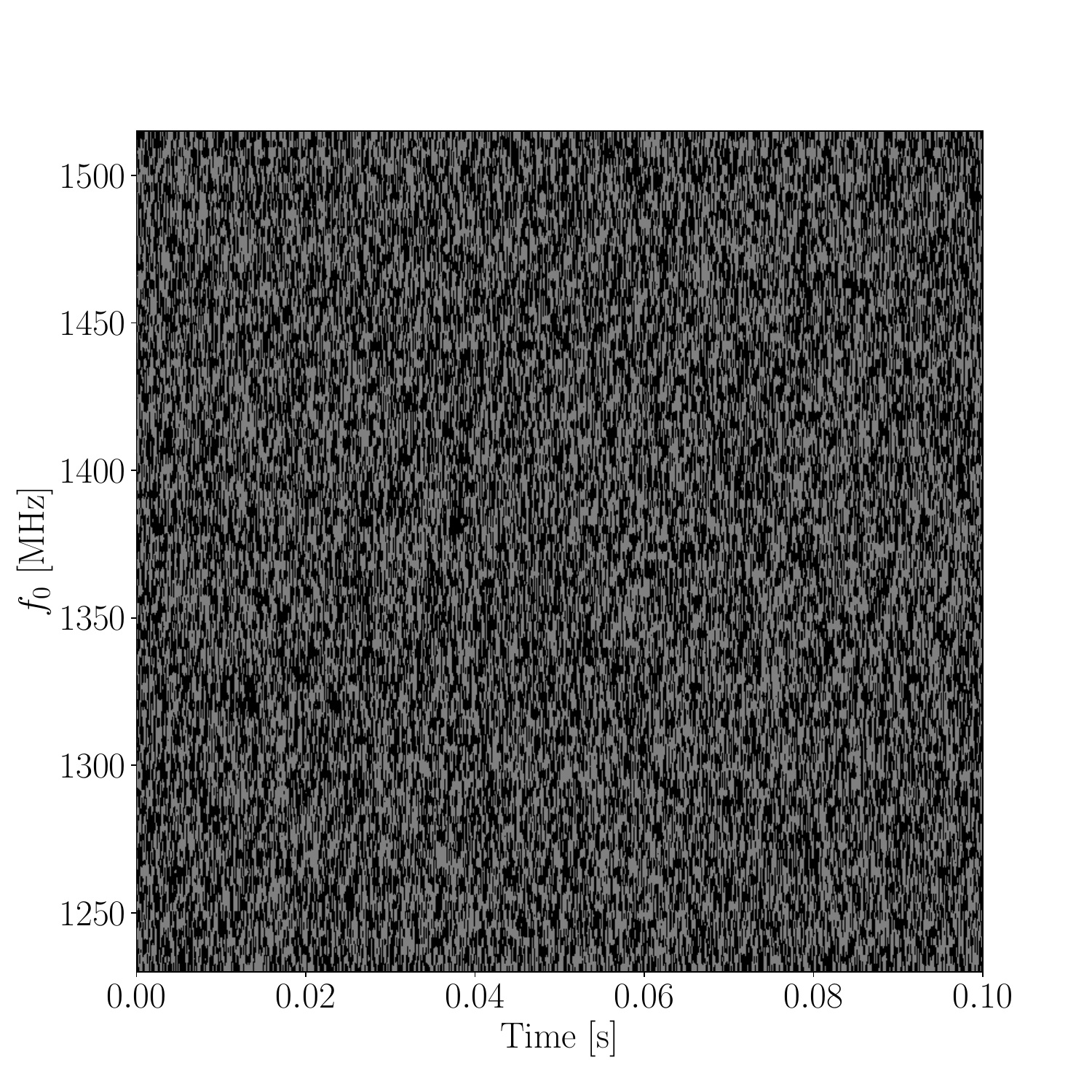}}
    \caption{Synthetic Parkes ``Murriyang'' multibeam system search-mode data, generated using the \texttt{simulatesearch} software package. The data without RFI are 1-bit digitized and constructed from Gaussian radiometer data $n(t)$ added to an injected pulsar signal $f_{\rm p}(t)$, simulated using the \texttt{simulateSystemNoise}, \texttt{simulateComplexPsr}, and \texttt{createSearchFile} subroutines, details of which are given in Section \ref{SubSec:SyntheticData}. The radio telescope and binary input parameters are reported in Table \ref{tab:SourceParameters}. The observing frequency $f_0$ (MHz) is reported on the vertical axis, centered at 1374 MHz \citep{Edwards_2001,Manchester_2001}. Time (units: seconds) is reported on the horizontal axis. A version of the diagram with RFI included appears in Figure \ref{Fig:specRFI}.}
    \label{Fig:specNoRFI}
\end{figure}

The 1-bit digitized PSRFITS search mode data output by \texttt{simulatesearch} are visualized in Figures \ref{Fig:specNoRFI} and \ref{Fig:specRFI}. The figures display the flux density as a pixellated greyscale frequency-time spectrogram, with time and frequency plotted on the horizontal and vertical axes, respectively.  Pulsar search mode data files record the flux density as a function of time and frequency channel. The data are 1-bit digitized, so the value of each time-frequency bin equals zero or one, reflected in the binary coloring (gray or black, respectively) in Figures \ref{Fig:specNoRFI} (noise plus signal) and \ref{Fig:specRFI} (noise, signal, and RFI, the latter appearing as a dark, black band).

\subsection{Idealizations}\label{SubSec:Idealizations}
We emphasize that the synthetic time series $x(t')$ generated by \texttt{tempo2}, \texttt{presto}, and \texttt{simulatesearch}, whose generation is discussed in the previous paragraphs, is highly idealized in important respects. For example, real radio survey data are affected by low-frequency measurement noise processes such as slowly varying instrumental gain fluctuations and telescope-pointing jitter, among others \citep{Lazarus_2015,VanHeerdan_2017,Zhang_2021,Luo_2022}. In practice, low-frequency noise manifests as an excess of power in the lower part of the Fourier spectrum, e.g.\ $\lesssim 10$ Hz for the Giant Meterwave Radio
Telescope Southern Sky Survey \citep{Singh_2022}. It presents challenges for detecting pulsars whose pulse frequencies satisfy $f_{\rm p}(t) \lesssim 50 \, \rm{Hz}$. Importantly, several spectrum whitening techniques \citep{Lorimer_2005} are implemented in modern pulsar search software packages, e.g.\ \texttt{presto} \citep{Ranson_2011} and \texttt{sigproc} \citep{Lorimer_2001}, mitigating the effects of low-frequency noise; see Sections 2.4.1 and 2.4.2 of \cite{VanHeerdan_2017} for overviews of the \texttt{sigproc} and \texttt{presto} spectrum whitening algorithms, respectively. Out of the $155$ compact binaries discovered to date  with $0.05 \lesssim P_{{\rm{b}}}/(1 \, {\rm{day}}) \lesssim 1$, five satisfy $f_{\rm{p,inj}} \lesssim 10 \, \rm{Hz}$, with the remaining $150$ systems satisfying $10 \lesssim f_{\rm{p,inj}}/(1 \, \rm{Hz}) \lesssim 700$. The HMM scheme in this paper is conceived primarily as a new way to discover compact, millisecond pulsars, focusing on the regime $P_{\rm{b}} < 1 \, \rm{day}$ and $f_{\rm{p,inj}} > 100 \, \rm{Hz}$, where it is reasonable to approximate the noise $n(t)$ as Gaussian. This is a starting point only; it can be generalized in future applications using (for example) \texttt{simulatesearch}. As just one example, Figure 2 of \cite{Luo_2022} displays the power spectrum of Gaussian radiometer noise supplemented with low-frequency red noise output by \texttt{simulatesearch}. 

A second idealization in the validation tests in Section \ref{Sec:Validation} and the performance tests in Section \ref{Sec:PerformanceTests} is to assume that RFI is excised from the synthetic radio survey data output by \texttt{simulatesearch} and \texttt{tempo2}. Although RFI is ubiquitous in real radio survey data, RFI-mitigation algorithms such as time-domain clipping and frequency-domain masking \citep{Lorimer_2005} are implemented in standard pulsar search software, e.g.\ \texttt{tempo2}'s \texttt{rfifind} \citep{Edwards_2006,Hobbs_2006}. In practice, real pulsar search pipelines mitigate the effects of RFI across several complex data processing stages, whose implementation lies outside the scope of this paper; see Figure 2 of \cite{VanHeerdan_2017} for details of the typical pulsar search data processing stages as well as \cite{Knispel_2013}, \cite{Ng_2015}, and \cite{Sobey_2022} for examples of RFI-mitigation strategies employed in real pulsar searches. As a rudimentary starting point, a worked example of binary pulsar detection with narrowband, impulsive RFI, injected using \texttt{simulatesearch}'s \texttt{simulateRFI} subroutine, is presented in Appendix \ref{App:HMMRFI} for the convenience of the reader. A fuller study of the HMM response to RFI is postponed to future work.

\begin{table}
\begin{center}
\begin{tabular}{cccc}
\hline
Quantity & Value (Section \ref{Sec:Validation}) & Range (Section \ref{Sec:PerformanceTests}) & Units \\ \hline
$f_{\rm{p,inj}}$ & 225.02 & -- & Hz  \\
$P_{\rm{b}}$ & 0.037 & $(0.012, 0.065)$ & days  \\
$a \sin i/c$ & $1.0 \times 10^{-2}$ & -- & lt-s \\
$e_{\rm{b}}$ & $6.0 \times 10^{-4}$ & -- & --  \\
$S$ & 1.0 & ($10^{-2}, 10^1$) & mJy  \\
$W_{50}$ & 0.15 & -- & -- \\
\hline
$T_{\rm{sys}}$ & 21 & -- & K \\
$G_{\rm{sys}}$ & 0.64 & -- & $\rm{K \, Jy^{-1}}$ \\
$n_{\rm{p}}$ & 2 & -- & -- \\
$\Delta f_{\rm{sys}}$ & 288 & -- & MHz \\
$t_{\rm{samp}}$ & 125 & -- & $\mu\rm{s}$ \\
$T_{\rm{obs}}$ & $10^4$ & -- & s \\
\hline
$B_{\rm min}$ & 50 & -- &  Hz\\
$B_{\rm max}$ & 1050 & -- & Hz \\
$\Delta f_{B}$ & 10 & -- & Hz \\
$\Delta f_{W}$ & 0.0032 & -- & Hz\\
$N_{B}$ & 100 & -- & -- \\
$T_{\rm coh}$ & 312 & -- & s \\
$N_{T}$ & 32 & -- &  --\\
\hline
\end{tabular}
\end{center}
\caption{Injected parameters of the representative test source in Section \ref{SubSec:TestSource} for the validation and performance tests in Sections \ref{Sec:Validation} and \ref{Sec:PerformanceTests}. The top section contains the signal, rotational, and orbital parameters of PSR J1953$+$1844, the pulsar with the shortest known orbital period. The middle section contains the synthetic Parkes ``Murriyang'' multibeam system parameters. The bottom section contains the HMM analysis parameters. }
\label{tab:SourceParameters}
\end{table}

\subsection{Noise-only response: setting a detection threshold}\label{SubSec:NoiseOnly}

We start by assessing the HMM response to the Gaussian radiometer noise $n(t)$ in the absence of a signal. The aim is to calculate a likelihood threshold $\mathcal{L}_{\rm th}$ for identifying pulsar candidates as a function of the subband false alarm probability $\alpha'$ according to the steps outlined in Section \ref{SubSec:Threshold}. Specifically, we generate $N_{\rm real} = 10^4$ Monte-Carlo realizations of a $N_T \times N_Q = 32 \times 3120$ spectrogram whose elements $P_{m,k}$ are exponentially distributed, with probability density function $p(P_{m,k}) = \exp(-P_{m,k})$. We construct $p(\mathcal{L})$ from the  $N_{\rm real} \times N_{Q}$ paths returned by the Viterbi algorithm (see Appendix \ref{App:Viterbi}) and set the likelihood cut-off for $\mathcal{L}_{\rm{tail}}$ to start at the $99.99$th percentile of $p(\mathcal{L})$. In this paper, we tolerate $\alpha' = 0.1$. Hence for the $N_{B} =100$ subbands analyzed in Section \ref{SubSec:FrequencyTracking}, we expect $\sim 10$ pulsar candidates to rise above $\mathcal{L}_{\rm th}$ due to random, Gaussian fluctuations. Solving Equations (\ref{Eq:MLELambda}) and (\ref{Eq:L_th}), we estimate $\Tilde{\lambda} = 0.51$ and $\mathcal{L}_{\rm th} = 96.2 \pm 0.052$, respectively. The reader is reminded that the uncertainty associated with $\mathcal{L}_{\rm th}$ is given by the square root of Equation (\ref{Eq:LthVariance}). Looking ahead, the results in Section \ref{SubSec:FrequencyTracking} yield six pulsar candidates above $\mathcal{L}_{\rm th}$, broadly consistent with the expectation above.

\begin{figure}
\centering{
    \includegraphics[width=0.75\textwidth, keepaspectratio]{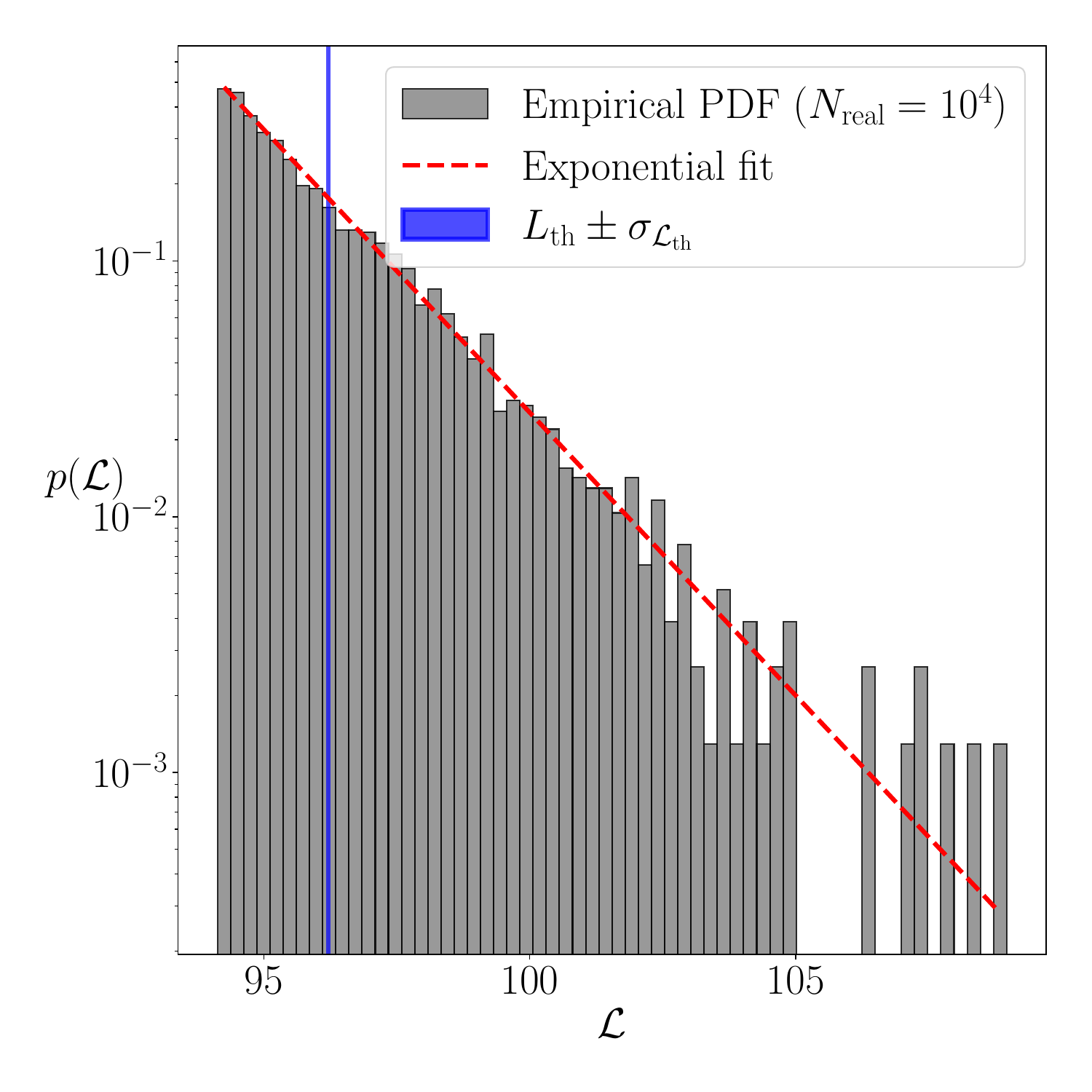}}
    \caption{Tail of the noise-only PDF $p(\mathcal{L}$) for $\mathcal{L}> \mathcal{L}_{\rm tail}$, used to set the detection threshold ${\cal L}_{\rm th}$ as a function of the per-subband false alarm probability $\alpha'$ . The gray histogram corresponds to the empirical PDF $p(\mathcal{L}$) for $\mathcal{L}> \mathcal{L}_{\rm tail}$ computed by analyzing $N_{\rm real} = 10^4$ Monte-Carlo realizations of a 10 Hz subband with $N_T = 32$ time bins and $N_Q = 3120$ frequency bins using the Viterbi algorithm in Appendix \ref{App:Viterbi}. Overplotted as a red, dashed line is Equation (\ref{Eq:TailDist}) for $k = 10^{-4}$ and $\Tilde{\lambda} = 0.51$. The likelihood threshold $\mathcal{L}_{\rm th} = 96.2\pm 0.052$, calculated in Section \ref{SubSec:NoiseOnly}, is overplotted as a blue, shaded, vertical region. }
    \label{Fig:likelihood}
\end{figure}

In Figure \ref{Fig:likelihood} we plot $p(\mathcal{L})$ as a gray histogram for $\mathcal{L}>\mathcal{L}_{\rm tail}$. The histogram counts the $\mathcal{L}>\mathcal{L}_{\rm tail}$ outliers associated with the $N_{\rm real}\times N_{Q}$ log-likelihoods $\mathcal{L}$ returned by the Viterbi algorithm.  Overplotted are the likelihood threshold $\mathcal{L}_{\rm th} = 96.2\pm0.052$ (blue shaded region), estimated using Equation (\ref{Eq:L_th}), and an empirical fit to the tail of $p(\mathcal{L})$ (dashed red line), estimated using Equation (\ref{Eq:TailDist}) with $\Tilde{\lambda} = 0.51$ and $k=10^{-4}$. At first glance it may appear that too many $\mathcal{L}$ samples satisfy $\mathcal{L} > \mathcal{L}_{\rm th}$. However we remind the reader that the gray histogram is constructed from $N_{\rm real} = 10^4$ Monte-Carlo realizations of a time-frequency spectrogram, so we expect $\sim 10^3$ false alarms for $\alpha' = 0.1$, as observed in Figure \ref{Fig:likelihood}.  

We remind the reader of an important difference between setting a likelihood threshold in tests with synthetic data and in real pulsar searches. In practice, one typically has $N_{\rm real} = 1$. The single realization is affected by radio frequency interference as well as low-frequency measurement noise processes such as slowly varying instrumental gain fluctuations and telescope-pointing jitter, among others \citep{Lazarus_2015,VanHeerdan_2017,Zhang_2021,Luo_2022}. One possible approach is to estimate the local noise background of a real search using (for example) off-source analysis, an approach popular with the continuous gravitational-wave community \citep{Astone_2014,Abbott_2021,Abbott_2022c,DOnofrio_2023}. Setting a detection threshold when using real radio survey data will be addressed in detail in a forthcoming manuscript.

\subsection{Noise plus signal: tracking the pulse frequency} \label{SubSec:FrequencyTracking}

In this section we analyze the synthetic pulsar survey data in Section \ref{SubSec:SyntheticData} to search for a synthetic source, namely the representative test source in Section \ref{SubSec:TestSource}, whose signal, rotational, and orbital parameters are reported in the top section of Table \ref{tab:SourceParameters}. The total search band $50 \leq f_0/(1 \, \rm{Hz}) \leq 1050$, reported in the bottom section of Table \ref{tab:SourceParameters}, is astrophysically motivated and justified as follows. The upper limit corresponds approximately to a pulsar's centrifugal break-up frequency \citep{Cook_1994,Lattimer_2007,Gittins_2024} and encompasses the pulse frequencies of all discovered millisecond pulsars, the maximum of which is 716 Hz \citep{Hessel_2006}. The lower limit marks approximately where Fourier-based techniques begin to lose sensitivity, e.g.\ due to the low-frequency measurement noise discussed in Section \ref{SubSec:SyntheticData}.
We divide the $B_{\rm{max}} - B_{\rm{\rm min}} = 1000$ Hz search band into $N_{B} = 100$ subbands of width $\Delta f_{B} = 10$ Hz. The associated frequency bin widths are $\Delta f_{W} = 0.0032 \, \rm{Hz}$. The $T_{\rm{obs}} = 10^4$ s observation comprises $N_{T} = 32$ coherent segments of duration  $T_{\rm{coh}} = 312$ s. We refer the reader to Section \ref{SubSec:Tcoh} and Appendix \ref{App:TimeBins} for further details.

Following the data reduction process in Section \ref{SubSec:DataProducts}, the HMM is solved recursively via the classic Viterbi algorithm \citep{Viterbi_1967}, whose pseudocode is summarized in Appendix \ref{App:Viterbi}. That is, for each coherent data segment of duration $T_{\rm{coh}}$, we apply Equation (\ref{Eq:Power}) to the $N_{Q}$ frequency bins and estimate $\mathcal{L} = \ln P(Q|O)$ over the full observation interval $T_{\rm{obs}}$ by evaluating the logarithm of the product on the right-hand side of Equation (\ref{Eq:MaxPathQ}). For each subband, the Viterbi algorithm yields $Q^*$, i.e. the hidden state sequence that maximizes $P(Q|O)$, where $Q^*$ is calculated according to Equation (\ref{Eq:MaxPathQ}) and Steps \ref{Alg:Backtrack1}--\ref{Alg:Backtrack3} of Algorithm \ref{Alg:Viterbi}. 

In its simplest form, the output of a real radio pulsar survey is a list of pulsar candidates requiring follow-up analysis, e.g.\ to distinguish terrestrial RFI from real astrophysical sources. The final stages of candidate selection involve visual inspection of pulsar candidates, which have been filtered using (for example) graphical interfaces, e.g.\ \texttt{reaper} \citep{Faulkner_2004} and \texttt{jreaper} \citep{Keith_2009}, heuristic candidate ranking strategies \citep{Lee_2013,Clark_2017,Patel_2018}, machine learning techniques \citep{Bethapudi_2018,Sanidas_2019,Bhat_2023}, and so on \citep{Balakrishnan_2021}. 

\begin{figure*}
\centering{
    \includegraphics[width=\textwidth, keepaspectratio]{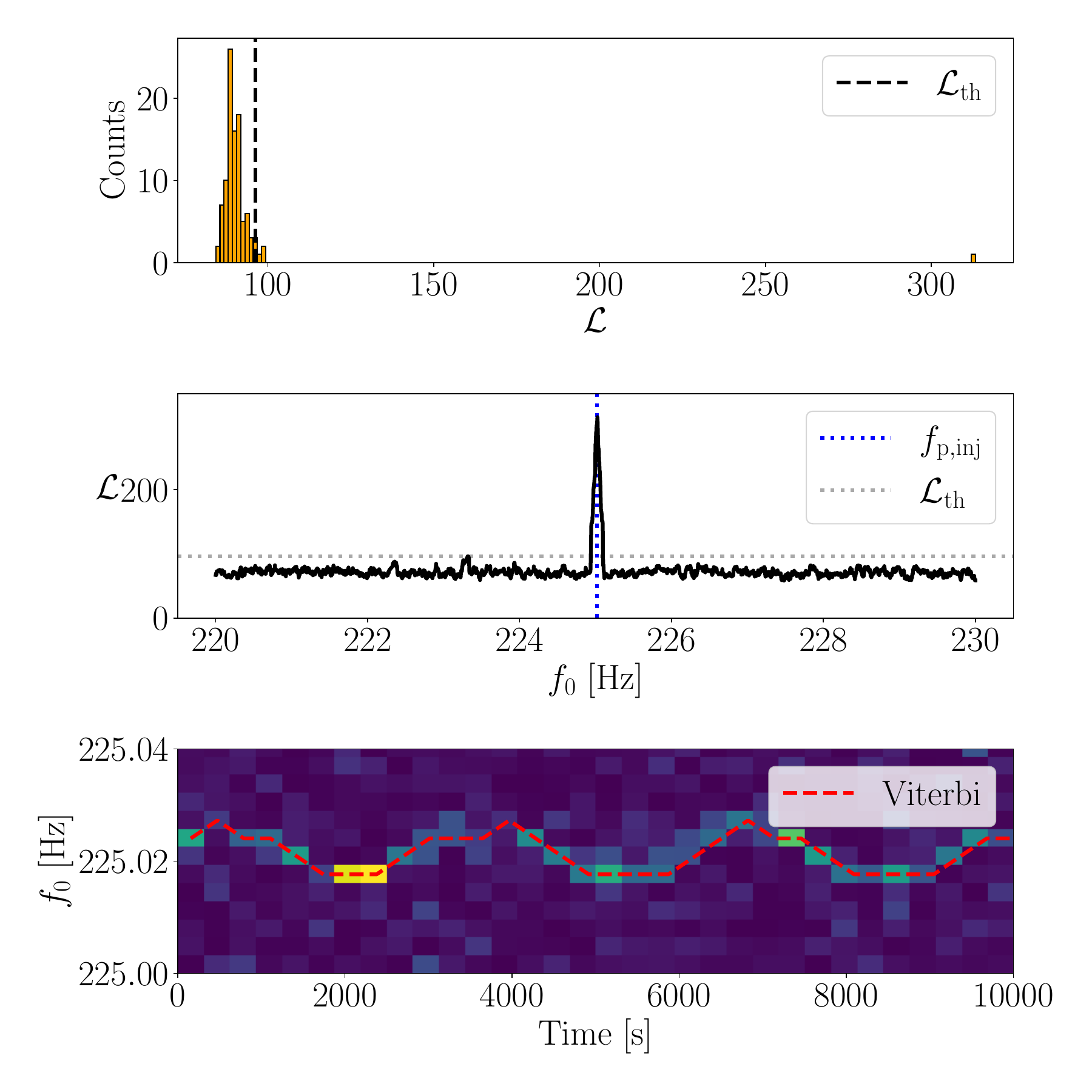}}
    \caption{Frequency tracking results for the representative test source in Section \ref{Sec:Validation} across the full search band $50 \leq f_0/ (1\, {\rm Hz}) \leq 1050$ (top panel) and in a single subband $220 \leq f_0 / (1\, {\rm Hz}) \leq 230$ which contains an above-threshold outlier with ${\cal L} > {\cal L}_{\rm th}$ (middle and bottom panels). (Top panel.) Histogram of the $100$ $\mathcal{L}$ values returned by the Viterbi algorithm for the maximal paths in the $N_B = 100$ subbands for the synthetic noise plus signal validation test in Section \ref{SubSec:FrequencyTracking}. Six out of the 100 maximal paths are above-threshold outliers. The black, dashed, vertical line indicates the log-likelihood threshold $\mathcal{L}_{\rm th}$ set in Section \ref{SubSec:Threshold}.  (Middle panel.) Log-likelihood $\mathcal{L} = \ln P(Q^*|O)$ of the Viterbi paths ending in 3120 frequency bins versus the terminating bin frequency $f_0$ (units: Hz), plotted as a black curve. The blue, dotted, vertical line indicates the injected pulse frequency $f_{\rm p, inj} = 225.02 \, \rm{Hz}$, reported in the top section of Table \ref{tab:SourceParameters}. The gray, dotted, horizontal line corresponds to the likelihood threshold $\mathcal{L}_{\rm th} = 96.2 \pm 0.052$, calculated in Section \ref{SubSec:NoiseOnly}.  (Bottom panel.) Magnified subset of the frequency-time spectrogram with 32$\times$12 pixels whose coloring indicates in a heat map the value of the normalized Fourier power, calculated according to Equation (\ref{Eq:Power}). The red, dashed curve is the optimal hidden state sequence $f_{\rm p}(t)$ output by the Viterbi algorithm.}
    \label{Fig:NoRFIPlots}
\end{figure*}

In the top panel of Figure \ref{Fig:NoRFIPlots}, we plot a histogram of the $100$ ${\cal L}$ values returned by the Viterbi algorithm for the maximal paths in the $N_B=100$ subbands for the synthetic noise plus signal validation test analyzed above. Six of the maximal paths return ${\cal L} > {\cal L}_{\rm th} = 96.2\pm0.052$, with $\max {\cal L} = 345.18$ and $\min {\cal L} = 84.27$ among the six above-threshold outliers. Specifically, the Viterbi algorithm returns $N_Q$ possible state sequences $Q^\star = \{q^\star(t_0), \hdots, q^\star(t_{N_T}) \}$ in each of the $N_B = 100$ subbands (each 10 Hz wide) between $50 \leq f_{0}/(1 \, \rm{Hz}) \leq 1050$. That is, the analysis yields  $N_Q N_B = 3.120 \times 10^5$ possible state sequences to choose from. The top panel of Figure \ref{Fig:NoRFIPlots} displays the maximal path $Q^\star(O)$ in each of the $N_B = 100$ subbands, i.e.\ one maximal path $Q^\star(O)$ per subband, over the full search band $50 \leq f_{0}/(1 \, \rm{Hz}) \leq 1050$. In the rest of this section, we restrict attention to the $220 \leq f_0/(1 \, {\rm Hz}) \leq 230$ subband, i.e.\ we do not perform follow-up analysis on the remaining six potential candidates to avoid repetition. 

In the rest of Figure \ref{Fig:NoRFIPlots} we present the Viterbi frequency tracking results for the $220 \leq f_0/(1 \, {\rm Hz}) \leq 230$ subband. In the middle panel, we plot $\mathcal{L} = \ln P(Q^*|O)$ versus the observed frequency $f_0$ as a black curve. Overplotted as a gray, horizontal, dashed line is the likelihood threshold $\mathcal{L}_{\rm th}$ calculated in Section \ref{SubSec:NoiseOnly}. The blue, vertical, dashed line corresponds to the injected pulse frequency $f_{\rm p, inj}$ in Table \ref{tab:SourceParameters}. In the bottom panel, we plot the $N_T \times N_Q = 32 \times 3120$ bins of the time-frequency spectrogram. For each bin in the time-frequency plane, the coloring indicates the value of the detection statistic, i.e. the normalized Fourier power, calculated according to Equation (\ref{Eq:Power}), with brighter colors indicating a higher value in the same fashion as Figure \ref{Fig:specNoRFI}. Overplotted as a dashed, red curve is the optimal hidden state sequence $f_{\rm p }(t)$, constructed according to Steps \ref{Alg:Backtrack1}--\ref{Alg:Backtrack3} in Algorithm \ref{Alg:Viterbi}.

The results in Figure \ref{Fig:NoRFIPlots} exhibit three key features. First, the log-likelihood of the optimal Viterbi path, $\mathcal{L} = \ln P(Q^*|O) = 345.18 > \mathcal{L}_{\rm th}$, peaks near the injected pulse frequency $f_{\rm p, inj}$ (dashed blue line), visible in the top panel of Figure \ref{Fig:NoRFIPlots}. Specifically,  the optimal Viterbi path extends from $225.0176 \, {\rm Hz}$ to $225.0272 \, {\rm Hz}$, while the injected frequency equals $225.02 \, {\rm Hz}$. Second, the optimal Viterbi track is a significant detection. For example,  the probability of ${\cal L} \geq 345.18$ occurring by chance is less than $ 1 \times 10^{-20}$ (see Figure \ref{Fig:likelihood}). Alternative significance metrics, such as the normalized Viterbi score used in some gravitational wave searches, lead to the same conclusion \citep{Abbott_2017,Sun_2018}. This is not surprising: the representative pulsar in Table \ref{tab:SourceParameters} is relatively bright, with flux density $S = 1.0 \, \rm{mJy}$. Third, the temporal evolution of $f_{\rm p}(t)$, inferred by the Viterbi algorithm in Appendix \ref{App:Viterbi} and plotted as a red, dashed curve in the bottom panel of Figure \ref{Fig:NoRFIPlots}, is broadly consistent with the expected pulse frequency modulation due to binary motion. Specifically, we observe quasisinusoidal pulse frequency modulations with zero-to-peak amplitude $\approx 7.25 \times 10^{-3} \, {\rm Hz}$ and period $\approx 3 \times 10^3 \, {\rm s}$ for $0 \leq t' \leq 10^4 \, {\rm s}$ in the bottom panel of Figure \ref{Fig:NoRFIPlots}. 

\section{Performance tests}\label{Sec:PerformanceTests}

The results in Figure \ref{Fig:NoRFIPlots} are encouraging. However, they refer to a single, random realization of the Gaussian radiometer data $n(t)$ in Section \ref{SubSec:SyntheticData} added to an injected pulsar signal $f_{\rm p}(t)$ for the representative test source in Section \ref{SubSec:TestSource}. How representative are the results, if we repeat the experiment for different realizations of $n(t)$, while injecting a range of pulsar and orbital parameter combinations? We turn now to answer these questions. In Section \ref{SubSec:Flux} we place a limit on the minimum  flux density $S_{\rm min}$ required for a detection with the HMM. In Section \ref{SubSec:OrbitalPeriod} we quantify the sensitivity of the HMM as a function of $0.012 \leq P_{\rm b}/(1\, \rm{day}) \leq 0.065$ as well as $N_T$ and hence $T_{\rm coh}$ (with $T_{\rm obs}$ fixed). 

The performance tests in Sections \ref{SubSec:Flux} and \ref{SubSec:OrbitalPeriod} follow the same procedure as the validation tests in Section \ref{Sec:Validation} with two minor exceptions. First, we reduce the number of simulated frequency channels in the synthetic pulsar survey data in Section \ref{SubSec:SyntheticData} from 96 to eight, to reduce the computational overhead associated with generating Monte-Carlo realizations of $n(t)$ using \texttt{simulatesearch}'s \texttt{simulateSystemNoise} subroutine. The reduction does not affect the accuracy of the HMM. Second, we restrict attention to the optimal Viterbi path $\mathcal{L} = \ln P(Q^*|O)$ in the $220 \leq f_0/(1 \, \rm{Hz}) \leq 230$ subband. That is, we do not report false alarms due to Gaussian fluctuations in other subbands, again to reduce the computational overhead without loss of generality. 

We remind the reader that the performance tests in Section \ref{Sec:PerformanceTests} are not exhaustive. A full battery of tests using synthetic pulsar survey data lies outside the scope of the present paper, whose primary goal is to introduce a new, semi-coherent detection tool for discovering compact, millisecond pulsars in the regime $P_{\rm{b}} < 1 \, \rm{day}$ and $f_{\rm{p,inj}} > 100 \, \rm{Hz}$. Preliminary experiments suggest that the HMM's performance depends weakly on $f_{\rm p, inj}$ and $e_{\rm b}$, but the results are not presented in this introductory paper for brevity. Likewise, the HMM's performance does not depend on the secular spin-down rate of the pulsar $\dot{f}_{\rm p, inj}$, e.g.\ due to electromagnetic braking, as confirmed in Appendix \ref{App:PulsarBrak}. 

\subsection{Minimum detectable flux density $S_{\rm min}$} \label{SubSec:Flux}

\begin{figure*}
\centering{
    \includegraphics[width=\textwidth, keepaspectratio]{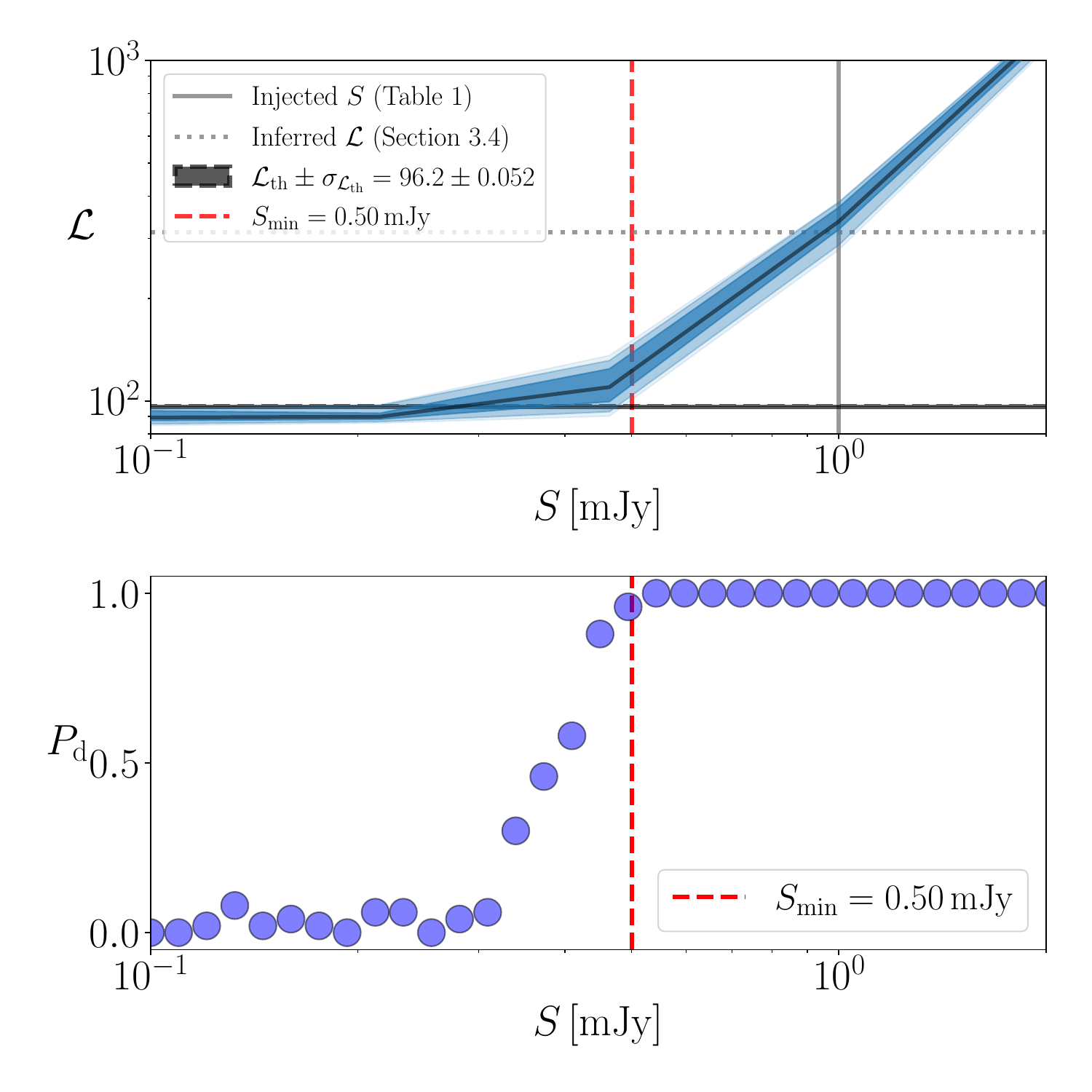}}
    \caption{Minimum detectable flux density $S_{\rm min}$ (units: mJy) for the representative test source in Section \ref{Sec:Validation} in a single subband $220 \leq f_0/(1 \, \rm{Hz}) \leq 230$. (Top panel.) Log-likelihood $\mathcal{L} = \ln P(Q^*|O)$ of the optimal Viterbi paths versus the injected flux density $10^{-1} \leq S/(1 \, \rm{mJy}) \leq 2.0$. For every injected $S$, the experiment is repeated using 50 Monte-Carlo realizations of $n(t)$. The median value of $\mathcal{L}$ as a black curve. The dark blue to light blue shaded regions correspond to the 68\%, 95\%, and 99\% credible intervals, respectively.  The black, dashed, horizontal line indicates the log-likelihood threshold set in Sections \ref{SubSec:Threshold} and \ref{SubSec:NoiseOnly}. The injected flux density $S = 1.0 \, \rm{mJy}$ (gray, solid line) as well as the inferred $\mathcal{L} = 345.18$ (gray, dotted line) from the validation tests in Section \ref{Sec:Validation} are overplotted by way of comparison. The red, dashed, vertical line indicates the inferred minimum detectable flux density $S_{\rm min} = 0.50 \, \rm{mJy}$. (Bottom panel.) Probability of detection $P_{\rm d}$ versus the injected flux density $10^{-1} \leq S/(1 \, \rm{mJy})  \leq 2.0$ using 50 Monte-Carlo realizations of $n(t)$. Every blue point corresponds to the number of realizations that exceed $\mathcal{L}_{\rm th}$ divided by the total number of Monte-Carlo realizations of $n(t)$. The red, dashed, vertical line indicates $S_{\rm min} = 0.50 \, \rm{mJy}$.}
    \label{Fig:MinFlux}
\end{figure*}

The brightness of a pulsar is a key factor controlling whether or not it is detectable. We quantify this in Figure \ref{Fig:MinFlux} by determining empirically the minimum detectable flux density $S_{\rm min}$, for the representative system in Table \ref{tab:SourceParameters}. Specifically, we assess the HMM's response to different injected values of flux density $S$ via Monte-Carlo simulations and quantify the reliability with which a detection is claimed using two metrics, namely the log-likelihood of the maximal Viterbi path $\mathcal{L} = \ln P(Q^*|O)$ and the probability of detection $P_{\rm d}$. 

In the top panel of Figure \ref{Fig:MinFlux}, we plot $\mathcal{L} = \ln P(Q^*|O)$ versus the injected flux density $S$. We vary $S$ in 10 evenly spaced logarithmic steps, with $10^{-2} \lesssim S/(1 \, \rm{mJy}) \lesssim 10^1$. For every $S$, the experiment is repeated 50 times. We plot the median value of $\mathcal{L}$ as a black curve as well as the corresponding 68\%, 95\%, and 99\% credible intervals, visible in the top panel of Figure \ref{Fig:MinFlux} as three, shaded blue regions. Overplotted as a black, horizontal, shaded region is the likelihood threshold $\mathcal{L}_{\rm th} = 96.2 \pm 0.052$ set in Section \ref{SubSec:NoiseOnly}. By way of comparison, we also plot the injected flux density $S = 1.0 \, \rm{mJy}$ (gray, solid line) and the inferred $\mathcal{L} = \ln P(Q^*|O) = 345.18 > \mathcal{L}_{\rm th}$ (gray, dotted line) from the validation tests in Section \ref{SubSec:FrequencyTracking} for the representative test source in Section \ref{SubSec:TestSource}. A red, dashed, vertical line indicates the minimum flux density $S_{\rm min} = 0.50 \, \rm{mJy}$ required for a detection. The horizontal and vertical axes are plotted on $\log_{10}$ scales.

In the bottom panel of Figure \ref{Fig:MinFlux}, we plot the probability of detection $P_{\rm d}$ as a function of the injected flux density $S$ as blue points. We vary $S$ in 50 evenly spaced steps, with $10^{-1} \leq S/(1 \, \rm{mJy}) \leq 10^{1}$. For every $S$, we repeat the experiment 50 times. Specifically, every blue point in the $P_{\rm d}$--$S$ plane corresponds to the number of realizations that exceed $\mathcal{L}_{\rm th}$ divided by the total number of realizations. A red, dashed, vertical line indicates $S_{\rm min} = 0.50 \, \rm{mJy}$. The horizontal and vertical axes are plotted on linear scales. 

The results in Figure \ref{Fig:MinFlux} exhibit three key features. First, for $S \lesssim 0.20 \, \rm{mJy}$, the results are broadly consistent with zero detections. In the top panel of Figure \ref{Fig:MinFlux}, $\mathcal{L} = \ln P(Q^\star|O)$  clusters near ${\cal L} \approx 90$. That is, we infer $\mathcal{L}<\mathcal{L}_{\rm th} = 96.2\pm 0.052$ in all realizations with $S \lesssim 0.20 \, \rm{mJy}$, except for three realizations associated with $S = 0.046$ mJy (with $\max \mathcal{L} = 99.17$), and three realizations associated with $S = 0.021$ mJy (with $\max \mathcal{L} = 97.20$). Second, we observe an approximately quadratic relationship of the form $\mathcal{L} = \ln P(Q^*|O) \propto S^{2}$ for $S \gtrsim 0.50 \, \rm{mJy}$, consistent with $\mathcal{L} = \ln P(Q^*|O) \propto P_{m,k} \propto S^2$. Third, for $S \geq 0.50 \, \rm{mJy}$, almost every realization qualifies as a potential detection, visible in the bottom panel of Figure \ref{Fig:MinFlux} as the horizontal sequence of blue points coincident with $P_{\rm d} = 1.0$. Hence, we infer $S_{\rm min} = 0.50 \, \rm{mJy}$. We remind the reader that the results in Figure \ref{Fig:MinFlux} are specific to the representative test source in Section \ref{SubSec:TestSource}, whose signal, rotational, and binary parameters are specified in the top section of Table \ref{tab:SourceParameters}. In principle, $S_{\rm min}$ depends on other factors such as $P_{\rm b}$, $f_{\rm p, inj}$, $W_{50}$, and $T_{\rm obs}$ \citep{Lorimer_2005,Lorimer_2008}. However, we find that the dependencies are weak, except for $P_{\rm b}$, which is discussed in Section \ref{SubSec:OrbitalPeriod}.

\subsection{Orbital period $P_{\rm b}$}\label{SubSec:OrbitalPeriod}

\begin{figure*}
\centering{
\includegraphics[width=0.75\textwidth, keepaspectratio]{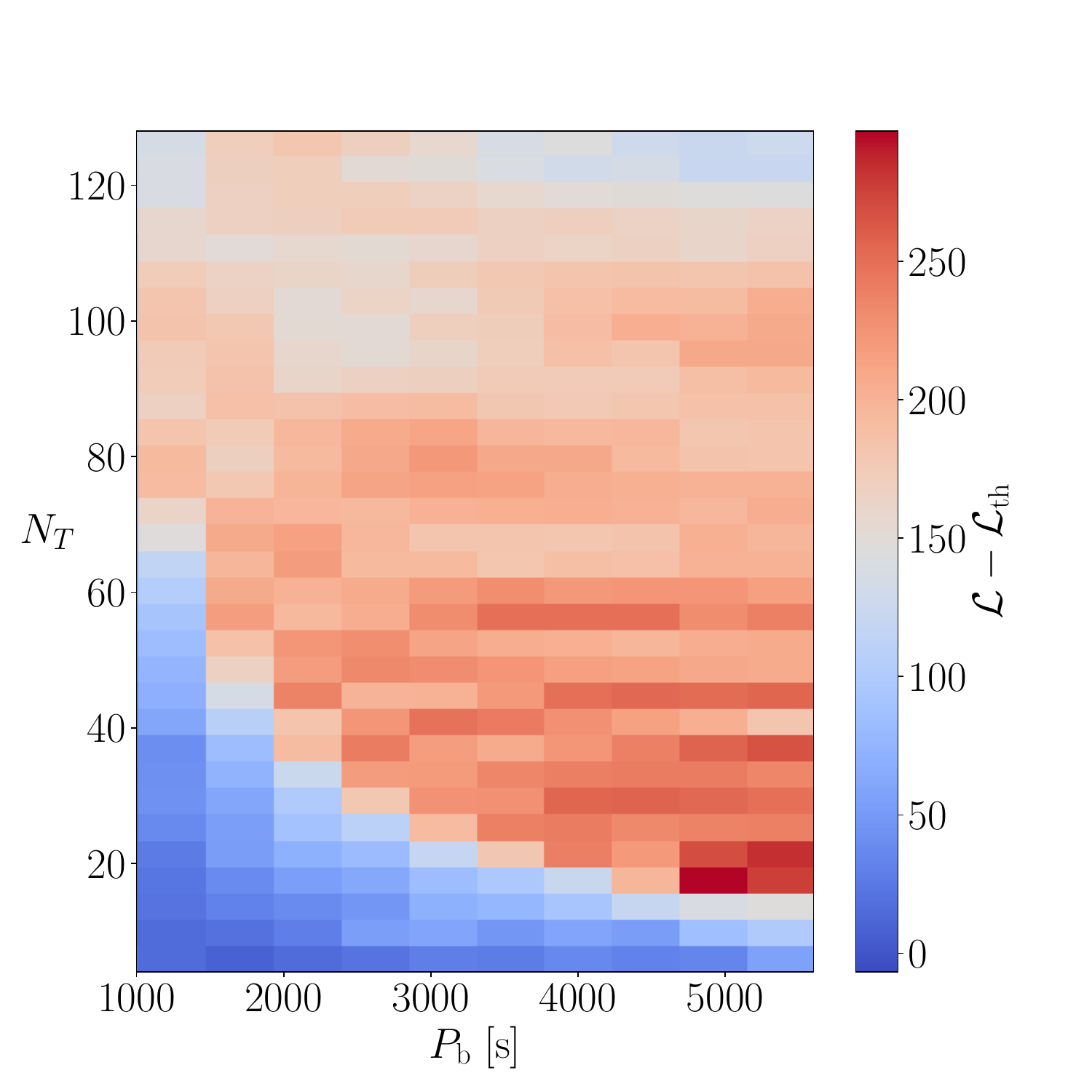}}
    \caption{Log-likelihood detection surface $\mathcal{L} - \mathcal{L}_{\rm{th}}$ ($\mathcal{L}$ minus $\mathcal{L}_{\rm th}$) versus the number of discrete time bins $4 \leq N_T \leq 128$ and the orbital period $1.0 \leq P_{\rm b}/(10^3 \, \rm{s}) \leq 5.6$. The $\mathcal{L} - \mathcal{L}_{\rm th}$ surface is visualized in cross section using a traditional heat map, whose red and blue coloring indicates higher and lower values of $\mathcal{L} - \mathcal{L}_{\rm th}$, respectively. For each pixel in the heat map, the reported value is calculated according to Equation (\ref{Eq:AverL}) and averaged over $N=50$ Monte-Carlo realizations, as discussed in Section \ref{SubSec:OrbitalPeriod}.}
    \label{Fig:Porb}
\end{figure*}

As the orbital period $P_{\rm b}$ decreases, the number of discrete time bins $N_T$ required for a detection increases. This makes sense physically: as $P_{\rm b}$ decreases, the Doppler modulation of $f_{\rm p}(t)$ due to the binary motion increases. Accordingly, the HMM requires more discrete time bins $N_T$, and hence shorter coherent segments $T_{\rm coh} = T_{\rm obs}/N_T$, to track the associated quasisinusoidal oscillations for constant $T_{\rm obs}$ (see Figure \ref{Fig:NoRFIPlots}). In Figure \ref{Fig:Porb} we quantify the sensitivity of the HMM as a function of $P_{\rm b}$ as well as the number of discrete time bins for $1.0 \leq P_{\rm b}/(10^3 \, \rm{s}) \leq 5.6$ and $4 \leq N_T \leq 128$. A recipe for selecting $N_T$ is given in Appendix \ref{App:TimeBins} \citep{Chandler_2003}. 

In Figure \ref{Fig:Porb} we present a surface plot of $\mathcal{L} - \mathcal{L}_{\rm th}$ ($\mathcal{L}$ minus $\mathcal{L}_{\rm th}$) as a function of $N_T$ and $P_{\rm b}$, visualized in cross section using a traditional heat map. For each $N_T$, we calculate $\mathcal{L}_{\rm th}$ according to the steps in Sections \ref{SubSec:Threshold} and \ref{SubSec:NoiseOnly}. That is, the likelihood threshold $\mathcal{L}_{\rm th} = \mathcal{L}_{\rm th}(N_T)$ is an explicit function of the number of discrete time bins $N_T$. We vary $N_T$ and $P_{\rm b}$ in 32 and 10 evenly spaced steps, with $4 \leq N_T \leq 128$ and $1000 \leq P_{\rm b}/(1 \, \rm{s}) \leq 5600$, respectively. Every $N_T$ and $P_{\rm b}$ pair yields one $\mathcal{L}$ value, corresponding to a single realisation of $n(t)$. Hence the analysis yields 320 $\mathcal{L}$ values in total. We repeat this experiment 50 times. The $\cal{L} - \cal{L}_{\rm th}$ values plotted in Figure \ref{Fig:Porb} are averaged over the 50 foregoing experiments. Specifically, we calculate $\mathcal{L} - \mathcal{L}_{\rm th}$ for the $i$-th $N_T$ bin and the $j$-th $P_{\rm b}$ bin in Figure \ref{Fig:Porb} according to
\begin{equation}\label{Eq:AverL}
(\mathcal{L} - \mathcal{L}_{\rm{th}})_{i,j} = N^{-1} \sum_{k=1}^N [\mathcal{L}_{k}(N_{T,i}, P_{{\rm{b}},j}) - \mathcal{L}_{\rm{th}}(N_{T,i})],
\end{equation}
where $\mathcal{L}_{k}$ denotes the log-likelihood associated with the $k$-th experiment for $N_{T,i}$ and $P_{{\rm b},j}$, for  $N=50$, $1 \leq i \leq 32$ and $ 1 \leq j \leq 10$. The red and blue coloring indicates higher and lower values of $\mathcal{L} - \mathcal{L}_{\rm th}$, respectively. The horizontal and vertical axes are plotted on linear scales. 

We draw the readers attention to two key features of Figure \ref{Fig:Porb}. First, there is a clear peak in the log-likelihood detection surface $\mathcal{L} - \mathcal{L}_{\rm th}$, visible as the group of dark red  pixels for $ P_{\rm b} \gtrsim 4.5 \times 10^3 \, \rm{s}$ and $16 \lesssim N_T \lesssim 36$. That is, for the representative test source in Table \ref{tab:SourceParameters}, the HMM is most sensitive in the foregoing region of the $N_T$--$P_{\rm b}$ parameter space. Second, we observe a transition from high $\mathcal{L} - \mathcal{L}_{\rm th}$ values (red coloring) to low $\mathcal{L} - \mathcal{L}_{\rm th}$ values (blue coloring) for constant $N_T \leq 64$, as $P_{\rm b}$ decreases. For example, for $N_T = 24$ and $N_T=36$ we observe a transition from red to blue coloring for $P_{\rm b} < 3 \times 10^3 \, \rm{s}$ and $P_{\rm b} < 2 \times 10^3 \, \rm{s}$, respectively. That is, as the modulation due to the binary motion increases, the number of discrete time bins $N_T$ required for a detection increases, as expected. 

\section{Computational cost}\label{Sec:CompCost}
One encouraging feature of the HMM search scheme developed in this paper is its speed. The CPU run time $T_{\rm run}$ (units: $\rm{s}$) of the Viterbi algorithm scales as $T_{\rm run} \propto N_T N_Q \ln N_Q$ \citep{Rabiner_1989,Quinn_2001} by taking advantage of dynamic programming and binary tree maximization to prune suboptimal paths efficiently.  In this section we confirm that the theoretical $T_{\rm run}$ scaling is achieved approximately in practice in the HMM implementation developed here.\footnote{The preliminary run-time tests in this section were performed
with a 3.2 GHz Apple M1 Pro processor.}  

\begin{figure}
\centering{
    \includegraphics[width=0.75\textwidth, keepaspectratio]{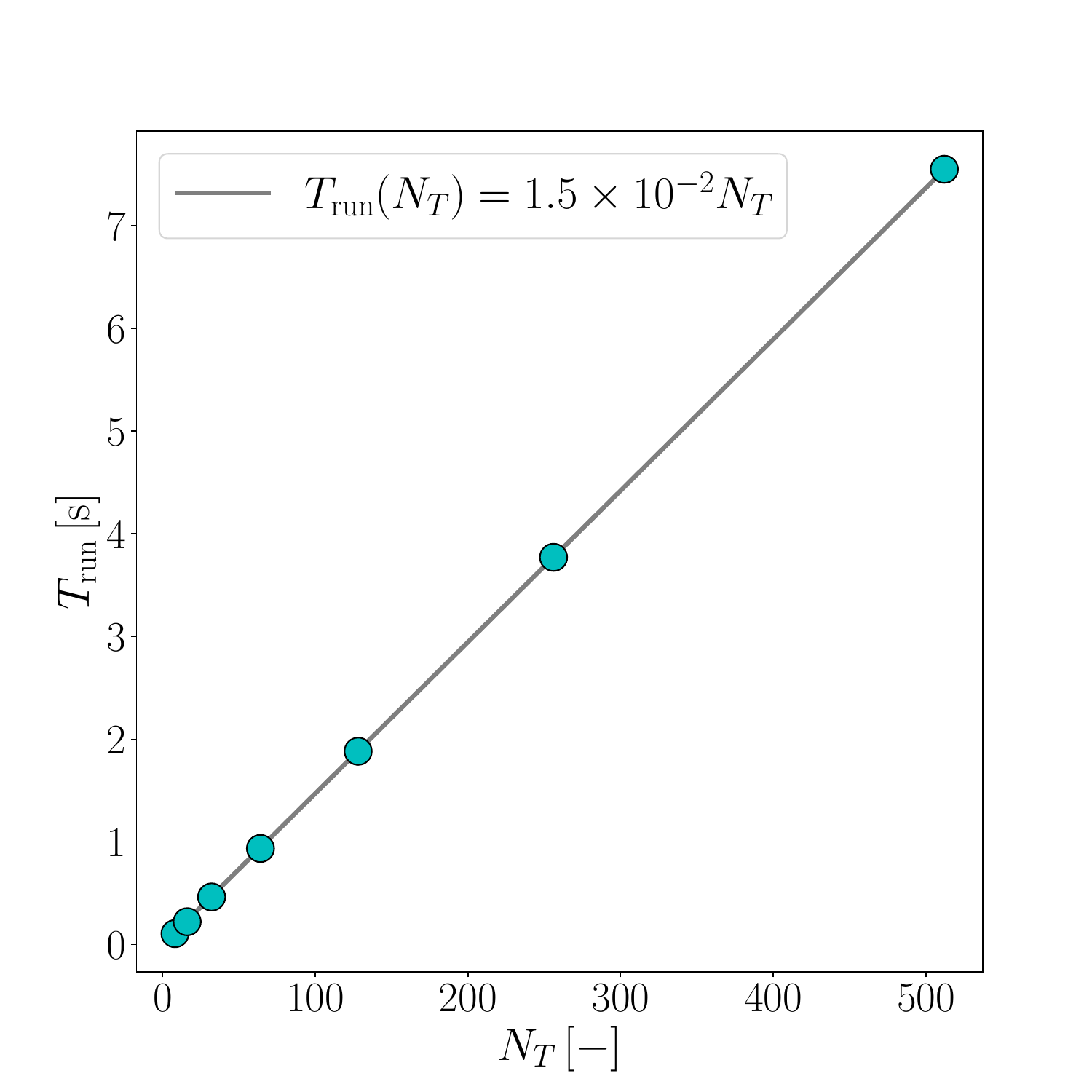}}
    \caption{Computational run-time tests of the Viterbi algorithm summarized in Appendix \ref{App:Viterbi}. We plot the per subband CPU run time $T_{\rm run}$ (units: seconds; cyan points) as a function of $N_T$, with $N_Q = 3120$ and $8 \leq N_T \leq 512$. Overplotted as a gray line is the equation $T_{\rm run} = 1.5 \times 10^{-2} N_T$, whose slope $1.5 \times 10^{-2}$ is determined empirically. }
    \label{Fig:RunTime}
\end{figure}

In Figure \ref{Fig:RunTime} we apply the Viterbi algorithm to the synthetic survey data generated in Section \ref{SubSec:SyntheticData} and plot the per subband CPU run time $T_{\rm{run}}$ as a function of $N_T$ using cyan points for $8 \leq N_T \leq 512$ and $N_Q = 3120$. Overplotted as a gray line is the equation $T_{\rm run} = 1.5 \times 10^{-2} N_T$, whose coefficient $1.5 \times 10^{-2}$ is determined empirically by fitting the slope. Therefore, the total CPU run time $T_{\rm{run.tot}}$ for a search involving $N_B$ subbands is given by
\begin{equation}\label{Eq:CPURunTime}
    T_{\rm run, tot} =  2.4 \times 10^{-1}  N_B (N_{T}/N_{T, \rm{ref}})  N_{Q} \ln N_{Q}/(N_{Q,\rm{ref}} \ln N_{Q,\rm{ref}}) \,\, \rm{s},
\end{equation}
where we adopt $N_{T,\rm{ref}} = 32$ and $N_{Q,\rm{ref}} = 3120$ as reference values. That is, for the validation tests in Section \ref{Sec:Validation}, the total CPU run time is $T_{\rm run,tot} \approx 24 \, \rm{s}$. Separate tests (not shown here for the sake of brevity) reveal that the $T_{\rm{run}}$ scaling depends weakly on the binary orbital eccentricity $e_{\rm b}$ and orbital period $P_{\rm b}$. Hence Equation (\ref{Eq:CPURunTime}) can be used by analysts to approximate the total CPU run time of future binary pulsar searches independent of $e_{\rm b}$ and $P_{\rm b}$, which are unknown at the outset in blind searches. 

It is challenging to compare directly Equation (\ref{Eq:CPURunTime}) with the total CPU run time of other, traditional pulsar search techniques, e.g.\ constant line-of-sight acceleration or jerk searches, because the underlying workflows are fundamentally different. For example, an acceleration search involves iterating over many trial acceleration values $a_1$, whereas the associated uncertainty in the secular orbital motion is handled by the transition matrix in Equation (\ref{Eq:TransSpec}) in this paper.  To convey the general flavor of such comparisons, however, consider as a representative example the acceleration search implemented by \cite{Balakrishnan_2022}. In the fourth column of Table 3 of the latter reference, the authors report the total CPU runtime $T_{\rm run, tot}$ of an acceleration search for one simulated binary [with $0.25 \lesssim P_{\rm b}/(1 \, \rm{day}) \lesssim 0.50$ and $T_{\rm obs} = 4.3 \times 10^3 \, \rm{s}$] as  $T_{\rm run,tot} \approx 6.1 \times 10^2 \, \rm{s}$, to be compared with $T_{\rm run,tot} \approx 24 \, \rm{s}$ for the HMM in this paper with $T_{\rm obs} = 10^4 \, \rm{s}$.\footnote{The total GPU runtimes $T_{\rm run, tot}^{\rm GPU}$ of acceleration, jerk, and fully coherent pulsar searches are reported in the fourth column of Table 3 of \cite{Balakrishnan_2022}. The total CPU runtime of the acceleration search in Table 3 of \cite{Balakrishnan_2022} is 25.5 times slower than $T_{\rm run, tot}^{\rm GPU}$, i.e.\ $T_{\rm run, tot} \approx 25.5 \, T_{\rm run, tot}^{\rm GPU} \approx 6.1 \times 10^2 \, \rm{s}$. } A full battery of tests comparing the computational runtime of pulsar search techniques is outside the scope of the present paper, and the foregoing CPU runtime of \cite{Balakrishnan_2022} is mentioned for completeness only.

\section{Conclusion}\label{Sec:Conclusions}

In this paper we demonstrate a new method for discovering pulsars in compact binaries in high-time-resolution radio survey data. The problem is formulated as a HMM, a powerful, statistical framework for frequency tracking under low signal-to-noise conditions. Within the HMM framework, the hidden state, i.e.\ the pulse frequency $f_{\rm p}(t)$, is related probabilistically to a time-ordered sequence of observations, i.e.\ a dedispersed and barycentred radio flux time series, via a detection statistic, i.e.\ the Fourier power. The HMM is solved recursively using the classic Viterbi algorithm \citep{Viterbi_1967,Rabiner_1989} through dynamic programming to infer the optimal path $f_{\rm p}^*(t)$, obviating the need to search over orbital parameters. The method is validated on synthetic Gaussian radiometer data deliberately in order to quantify its performance systematically under controlled conditions in Sections \ref{SubSec:SyntheticData} and \ref{SubSec:Idealizations}. 

The HMM approach builds on related work on continuous gravitational-wave searches \citep{Suvorova_2016,Suvorova_2017,Abbott_2017,Sun_2019,Abbott_2019,Middleton_2020,Melatos_2021,Beniwal_2021}, and pulsar glitch detection \citep{Melatos_2020,Dunn_2022a,Dunn_2023}. We emphasize that it does not supersede traditional pulsar detection techniques \citep{Johnston_1991,Camilo_2000,Ransom_2003,Allen_2013,Eatough_2013,Knispel_2013,Balakrishnan_2022}. Rather, it is complementary. Its computational speed makes it practical to run an HMM search and a traditional search in tandem without creating a bottleneck, for example, or to run an HMM search in quick-look first-pass mode, to be followed by a traditional search. The HMM framework can also be extended straightforwardly to accommodate colored noise, non-Gaussian noise artifacts, and modified signal models. By way of illustration, a worked example of binary pulsar detection in the presence of narrowband, impulsive RFI is presented in Appendix \ref{App:HMMRFI}.  

The HMM scheme is tested on simulated radio survey data, generated synthetically using the \texttt{simulatesearch} software package \citep{Luo_2022}. We initially focus on a single representative test source whose signal, rotational, and orbital parameters emulate those of PSR J1953$+$1844, the millisecond pulsar with the shortest known orbital period $P_{\rm b}=0.037 \, {\rm days}$ \citep{Nan_2011,Jiang_2019,Pan_2023}. The results in Section \ref{Sec:Validation} reveal that the method successfully recovers the injected pulse frequency $f_{\rm p, inj} =225.02 \, \rm{Hz}$ with an associated log-likelihood $\mathcal{L} = \ln P(Q^*|O) = 345.18 > \mathcal{L}_{\rm th} = 96.2 \pm 0.052$ (false alarm probability $\alpha' = 0.1)$. The detection is significant; the probability of $\mathcal{L} = 345.18$ occurring by chance is less than $10^{-20}$. The detection is also accurate; we find $|f_{\rm p}^\ast(t) - f_{\rm p,inj} | \leq 7.2 \times 10^{-3} \, {\rm Hz}$ for all $t$.

The results from a preliminary exploration of a broader pulsar parameter domain are presented in Section \ref{Sec:PerformanceTests}. We focus on two key factors controlling whether or not a pulsar is detectable, namely the flux density $S$ and the binary orbital period $P_{\rm b}$. The analysis reveals that the HMM detects injected sources down to a flux density given by $S_{\rm min} = 0.50 \, \rm{mJy}$  under observing conditions typical of previous multibeam surveys with the Parkes ``Murriyang'' Telescope. The method is sensitive to an orbital regime, which is expensive to search with traditional methods, e.g.\ the HMM successfully recovers the injected pulse frequency for $P_{\rm b} \geq 0.012 \, \rm{days}$ with $N_{T} = 64$, and hence $T_{\rm coh} = T_{\rm obs}/N_T =  156 \, \rm{s}$. Out of the $\approx 3380$ discovered pulsars recorded in the ATNF database (see Footnote \ref{FN:ATNFCat}), approximately 5 \% have $0.012 \leq P_{\rm b}/(1 \, {\rm day}) \leq 1.0$, so the HMM promises to be a useful tool in the future. Less than 1\% of the objects in the ATNF database have $S < S_{\rm min}$ and are currently out of its reach. This is encouraging and motivates a fuller exploration of the pulsar parameter space as well as further improvements to the method --- future goals which are beyond the scope of the present paper. 

One promising aspect of the HMM scheme is its speed. The Viterbi algorithm in Appendix \ref{App:Viterbi} takes advantage of dynamic programming and binary tree maximization to prune suboptimal paths efficiently. The CPU run time $T_{\rm run}$ of the Viterbi algorithm scales as $T_{\rm run} \propto N_T N_Q \ln N_Q$ \citep{Rabiner_1989,Quinn_2001}. For example, the total CPU run time for the validation tests in Section \ref{Sec:Validation} is $T_{\rm run, tot} \approx 24 \, \rm{s}$, for $N_T = 32, N_Q = 3120,$ and $N_B = 100$. Nonetheless, despite its speed, we reiterate that the HMM does not supersede well-established methods for binary pulsar detection, e.g.\ time-domain resampling \citep{Camilo_2000,Eatough_2013}, constant acceleration and jerk searches \citep{Johnston_1991,Anderson_2018}, and fully coherent searches \citep{Balakrishnan_2022}. HMM-based methods offer a fast and flexible statistical framework for frequency tracking, where the frequency wanders either stochastically or deterministically, and should be deployed in tandem with modern pulsar discovery techniques and timing software, e.g.\ \texttt{presto} \citep{Ranson_2011} and \texttt{tempo2} \citep{Edwards_2006,Hobbs_2006}. 

The next step is to apply the HMM to real radio survey data in collaboration with the pulsar timing community. As just one example, it would be interesting by way of preparation to check whether the HMM can detect the source PSR J1727--2951 \citep{Cameron_2020}, a black-widow binary whose binary and signal parameters, $P_{\rm b} = 0.39 \, \rm{days}$ and $S = 0.514 \, \rm{mJy}$, are broadly consistent with the synthetic pulsars analyzed in this paper.  It would also be interesting to reanalyze recent searches for binaries in globular clusters, e.g.\ in M28 and Terzan 5 using MeerKAT \citep{Douglas_2022,Padmanabh_2024} or in eight southern clusters using the Giant Metrewave Radio Telescope \citep{Gautam_2022}.

\begin{acknowledgments}
The authors acknowledge a helpful discussion with Ryan Shannon regarding the \texttt{pfits} software package.\footnote{\href{https://bitbucket.csiro.au/projects/PSRSOFT/repos/pfits/browse}{https://bitbucket.csiro.au/projects/PSRSOFT/repos/pfits/browse}} This research was supported by the Australian Research Council Centre of Excellence for Gravitational Wave Discovery (OzGrav), grant number CE170100004. The analysis employed the OzSTAR supercomputer facility at Swinburne University of Technology. The OzSTAR program receives funding in part from the Astronomy National Collaborative Research Infrastructure Strategy allocation provided by the Australian Government.
\end{acknowledgments}

\newpage
\appendix

\section{Viterbi algorithm}\label{App:Viterbi}

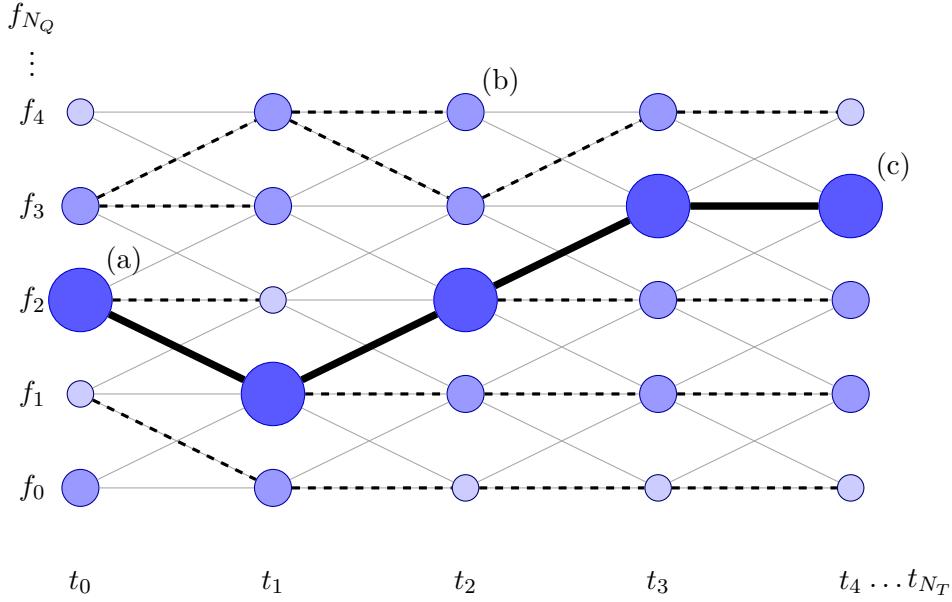
\begin{figure}
\centering
\begin{tikzpicture}[
    >=stealth,
    x=1.6cm,
    y=1.25cm,
    every node/.style={font=\small},
    stateLow/.style={circle, draw=blue!40!black, fill=blue!20,
                     minimum size=10pt, inner sep=0pt},
    stateMed/.style={circle, draw=blue!60!black, fill=blue!40,
                     minimum size=14pt, inner sep=0pt},
    stateHigh/.style={circle, draw=blue!80!black, fill=blue!65,
                      minimum size=24pt, inner sep=0pt},
    allconn/.style={draw=gray!70, line width=0.4pt},
    bestlocal/.style={draw=black, line width=1.2pt, dashed},
    bestglobal/.style={draw=black, line width=2.8pt}
]

\node at (0,   -1.0) {$t_0$};
\node at (1.6, -1.0) {$t_1$};
\node at (3.2, -1.0) {$t_2$};
\node at (4.8, -1.0) {$t_3$};
\node at (6.4, -1.0) {$t_4$};
\node at (6.9, -1.0) {$\hdots t_{N_{T}}$};

\node at (-0.4, 0.0) {$f_{0}$};
\node at (-0.4, 1.0) {$f_{1}$};
\node at (-0.4, 2.0) {$f_{2}$};
\node at (-0.4, 3.0) {$f_{3}$};
\node at (-0.4, 4.0) {$f_{4}$};
\node at (-0.4, 4.6) {$\vdots$};
\node at (-0.4, 5.0) {$f_{N_{Q}}$};


\node[stateMed]   (n00) at (0.0,0.0) {};
\node[stateLow]   (n01) at (0.0,1.0) {};
\node[stateHigh]  (n02) at (0.0,2.0) {}; 
\node[stateMed]   (n03) at (0.0,3.0) {};
\node[stateLow]   (n04) at (0.0,4.0) {};

\node[stateMed]   (n10) at (1.6,0.0) {};
\node[stateHigh]  (n11) at (1.6,1.0) {}; 
\node[stateLow]   (n12) at (1.6,2.0) {};
\node[stateMed]   (n13) at (1.6,3.0) {};
\node[stateMed]   (n14) at (1.6,4.0) {}; 

\node[stateLow]   (n20) at (3.2,0.0) {};
\node[stateMed]   (n21) at (3.2,1.0) {};
\node[stateHigh]  (n22) at (3.2,2.0) {}; 
\node[stateMed]   (n23) at (3.2,3.0) {};
\node[stateMed]   (n24) at (3.2,4.0) {}; 

\node[stateLow]   (n30) at (4.8,0.0) {};
\node[stateMed]   (n31) at (4.8,1.0) {};
\node[stateMed]   (n32) at (4.8,2.0) {};
\node[stateHigh]  (n33) at (4.8,3.0) {}; 
\node[stateMed]   (n34) at (4.8,4.0) {};

\node[stateLow]   (n40) at (6.4,0.0) {};
\node[stateMed]   (n41) at (6.4,1.0) {};
\node[stateMed]   (n42) at (6.4,2.0) {};
\node[stateHigh]  (n43) at (6.4,3.0) {}; 
\node[stateLow]   (n44) at (6.4,4.0) {};

\foreach \t/\tnext in {0/1,1/2,2/3,3/4}{
  \foreach \f in {0,...,4}{
    \foreach \df in {-1,0,1}{
      \pgfmathtruncatemacro{\fnext}{\f+\df}
      \ifnum\fnext>-1
        \ifnum\fnext<5
          \draw[allconn] (n\t\f) -- (n\tnext\fnext);
        \fi
      \fi
    }
  }
}

\draw[bestlocal] (n01) -- (n10); 
\draw[bestlocal] (n02) -- (n11); 
\draw[bestlocal] (n02) -- (n12); 
\draw[bestlocal] (n03) -- (n13); 
\draw[bestlocal] (n03) -- (n14); 

\draw[bestlocal] (n10) -- (n20);
\draw[bestlocal] (n11) -- (n21);
\draw[bestlocal] (n11) -- (n22); 
\draw[bestlocal] (n14) -- (n23);
\draw[bestlocal] (n14) -- (n24); 

\draw[bestlocal] (n20) -- (n30);
\draw[bestlocal] (n21) -- (n31);
\draw[bestlocal] (n22) -- (n32);
\draw[bestlocal] (n22) -- (n33); 
\draw[bestlocal] (n23) -- (n34);

\draw[bestlocal] (n30) -- (n40);
\draw[bestlocal] (n31) -- (n41);
\draw[bestlocal] (n32) -- (n42);
\draw[bestlocal] (n33) -- (n43); 
\draw[bestlocal] (n34) -- (n44);

\draw[bestglobal] (n02) -- (n11) -- (n22) -- (n33) -- (n43);

\node[above right=-0.15cm of n02] {(a)}; 
\node[above right=-0.15cm of n24] {(b)}; 
\node[above right=-0.15cm of n43] {(c)}; 

\end{tikzpicture}

\caption{Schematic diagram of dynamic programming and the classic Viterbi algorithm \citep{Bellman_1957,Viterbi_1967,Rabiner_1989,Quinn_2001}. The blue, circular nodes represent the $N_T \times N_Q$ elements in a time-frequency grid with discrete time  $t\in \{t_0, \hdots, t_{N_{T}}\}$ and frequency $f \in \{f_0, \hdots, f_{N_{Q}} \}$ bins along the horizontal (left to right) and vertical (bottom to top) directions, respectively. The size and coloring of each circular node in the $t$--$f$ grid represents the likelihood of a signal being present in that node, with larger, darker nodes corresponding to higher likelihood values. The $N_Q^{N_T}$ total possible paths are represented using gray lines. At time $t$, the Viterbi algorithm keeps track of the $N_Q$ optimal subpaths using dynamic programming, visible as dashed, black lines. In some cases, potential paths result in dead-ends, an example of which is visible at marker (b). At time $t = t_{N_{T}}$, the Viterbi algorithm selects the node with the highest likelihood value, depicted schematically as the largest and darkest terminating node, and calculated according to Equation (\ref{Eq:MaxPathQ}). The optimal Viterbi path is the path that leads to this node, visualized using the thick, black line between markers (a) and (c). The above schematic is based on Figure 4 of \cite{Gardner_2022}.}
\label{fig:viterbi_dp_single}
\end{figure}

The Viterbi algorithm, guided by Bellman's principle of optimality, scans the trellis of possible hidden state sequences and constructs $Q^*(O)$ through dynamic programming, depicted schematically in Figure \ref{fig:viterbi_dp_single} \citep{Bellman_1957,Rabiner_1989,Quinn_2001}.\footnote{Given an optimal hidden state sequence $Q^*$, Bellman's principle of optimality states that $Q^*$ does not change if one repeats the optimization procedure from a different starting state $q(t_1)_{\rm{new}} \in Q^*$. That is, the subpaths of $Q^*$ are optimal as well.} The logic and pseudocode of the Viterbi algorithm are summarized briefly in this appendix for the convenience of the reader. The summary follows closely the presentation in \cite{Melatos_2021}. 

The algorithm proceeds as follows. Let $\bm{\delta}(t_m)$ and $\bm{\Phi}(t_m)$ denote $N_Q$-dimensional vectors at time $t_m$ ($1 \leq m \leq N_T$) whose components are populated according to
\begin{equation}\label{Eq:Delta}
    \bm{\delta}_{q_{m'}}(t_m) = \max_{q_{m''}} {\rm{Pr}}[q(t_m) = q_{m'} | q(t_{m-1}) = q_{m''}; O^{(m)}],
\end{equation}
and
\begin{equation}
\bm{\Phi}_{q_{m'}}(t_m) = \argmax_{q_{m''}} {\rm{Pr}}[q(t_m) = q_{m'} | q(t_{m-1}) = q_{m''}; O^{(m)}],
\end{equation}
for $1 \leq m',m''\leq N_Q$, $O^{(m)} = \{ o(t_1), \hdots, o(t_m) \}$, and ${\rm{Pr}}[q(t_m) = q_{m'} | q(t_{m-1}) = q_{m''}; O^{(m)}] = L_{o(t_m) q_{m'}} A_{q_{m'} q_{m''}} \delta_{q_{m''}}(t_{m-1})$ \citep{Melatos_2021}. That is, $\bm{\delta}_{q_{m'}}(t_m)$ and $\bm{\Phi}_{q_{m'}}(t_m)$ store the $N_Q$ maximum transition probabilities at time $t = t_m$, and the hidden states at time $t = t_{m-1}$ that lead to the $N_Q$ maximum probabilities at time $t = t_m$, respectively. Specifically, for every step forward $m$ in the recursive propagation (Steps \ref{Alg:Step6}--\ref{Alg:Step10} in Algorithm \ref{Alg:Viterbi}), the Viterbi algorithm discards all but $N_{Q}$ possible paths, reducing the number of comparisons from $\propto N_Q^{N_T}$ using a brute-force approach to $\propto N_T N_Q \ln N_Q$ using dynamic programming and binary maximization \citep{Bellman_1957, Quinn_2001}. The optimal sequence of hidden states $Q^*(O)$, conditional on the form of the HMM, i.e.\ Equations (\ref{Eq:EmissionSpec})--(\ref{Eq:PriorSpec}), is constructed by backtracking through the $N_T$ vectors $\bm{\delta}(t_m)$ and $\bm{\Phi}(t_m)$ according to Steps \ref{Alg:Backtrack1}--\ref{Alg:Backtrack3} in Algorithm \ref{Alg:Viterbi}.

\newpage
\begin{aalgorithm}[Viterbi algorithm]\label{Alg:Viterbi}
\begin{algorithmic}[1]
\State Inputs: $\Pi$, $L$, $A$.  
\State Outputs: $Q^* = \{q^*(t_0),\hdots, q^*(t_{N_T})\}$
\State Initialization: \For{$1 \leq i \leq N_Q$}
\State $\delta_{q_i}(t_0) = L_{o(t_0)q_i}\Pi_{q_i}$
\EndFor
\State Recursion: \label{Alg:Step6}
\For{$1 \leq k \leq N_T$}
\For{$1 \leq i \leq N_Q$}
\State  $\delta_{q_i}(t_k) = L_{o(t_k) q_i} \displaystyle \max_{1\leq j \leq N_Q} [A_{q_iq_j}  \delta_{q_j}(t_{k-1})]$
\State $\Phi_{q_i}(t_k) = \displaystyle \argmax_{1 \leq j \leq N_Q} [A_{q_i q_j}\delta_{q_j}(t_{k-1})]$ \label{Alg:Step10}
\EndFor
\EndFor
\State Termination: 
\For{$1 \leq i \leq N_Q$}
\State $\max {\rm{Pr}}(Q|O) = \displaystyle \max_{q_i} \delta_{q_i}(t_{N_T})$ 
\State $q^*(t_{N_T}) = \displaystyle \argmax_{q_i} \delta_{q_i}(t_{N_T})$
\EndFor
\State Backtracking: \label{Alg:Backtrack1}
\For{$0 \leq k \leq N_T - 1$}
\State $q^*(t_k) = \Phi_{q^*(t_{k+1})}(t_{k+1})$ \label{Alg:Backtrack3}
\EndFor
\State \Return $Q^* = \{q^*(t_0),\hdots q^*(t_{N_T})\}$
\end{algorithmic}
\end{aalgorithm}

\newpage

\section{Selecting $N_T$}\label{App:TimeBins}
The number of discrete time bins $N_T$ is a key input into the HMM in Section \ref{Sec:HMMAlg} and should be set judiciously to: (i) maximize the single segment signal-to-noise ratio; and (ii) avoid $f_{\rm p}(t)$ wandering by more than one frequency bin between coherent segments. One possible recipe for selecting $N_T$ proceeds as follows \citep{Chandler_2003}. 

A pulsar whose frequency $f_{\rm p}(t)$ wanders harmonically and deterministically due to binary motion, remains within one frequency bin of width $N_T/T_{\rm obs}$ over an interval of duration $T_{\rm coh} = T_{\rm obs}/N_T$ provided that one has
\begin{equation} \label{Eq:NT_constraint}
|d f_{\rm p}/dt| (T_{\rm obs}/N_T) \leq N_T/T_{\rm obs}.
\end{equation}
For a pulsar whose orbit is nearly circular ($e_{\rm b} \approx 0$), the first time derivative of $f_{\rm p}(t)$ satisfies \citep{Chandler_2003,Lorimer_2005}
\begin{equation}\label{Eq:frequencyDer}
    \frac{d f_{\rm p}}{dt} \leq \frac{2 \pi f_{\rm p} \beta}{P_{\rm b}} \sin \left[ 2 \pi t/P_{\rm b} + \phi(t'_0) \right].
\end{equation}
In Equation (\ref{Eq:frequencyDer}), the orbital phase is denoted by $\phi$, and we remind the reader that the dimensionless parameter $\beta$ is given by $\beta = V_1 \sin i /c$. Upon substituting the right-hand-side of Equation (\ref{Eq:frequencyDer}) into Equation (\ref{Eq:NT_constraint}), we find \citep{Chandler_2003}
\begin{equation}\label{Eq:NQMin}
    N_T \geq (2 \pi f_{\rm p, inj} \, \beta \, T_{\rm obs}^2/P_{\rm b})^{1/2}.
\end{equation}
For non-overlapping coherent segments, $N_T$ is restricted to powers of two to maximize the speed of the Fourier transforms in Equation (\ref{Eq:FFT}) \citep{Chandler_2003}. In the validation and performance tests in Sections \ref{Sec:Validation}, we adopt $N_T = 32$ for the representative test source in Table \ref{tab:SourceParameters}; see Figure \ref{Fig:Porb} for details. 

\section{HMM response to RFI}\label{App:HMMRFI}
\begin{figure}
\centering{
    \includegraphics[width=0.75\columnwidth, keepaspectratio]{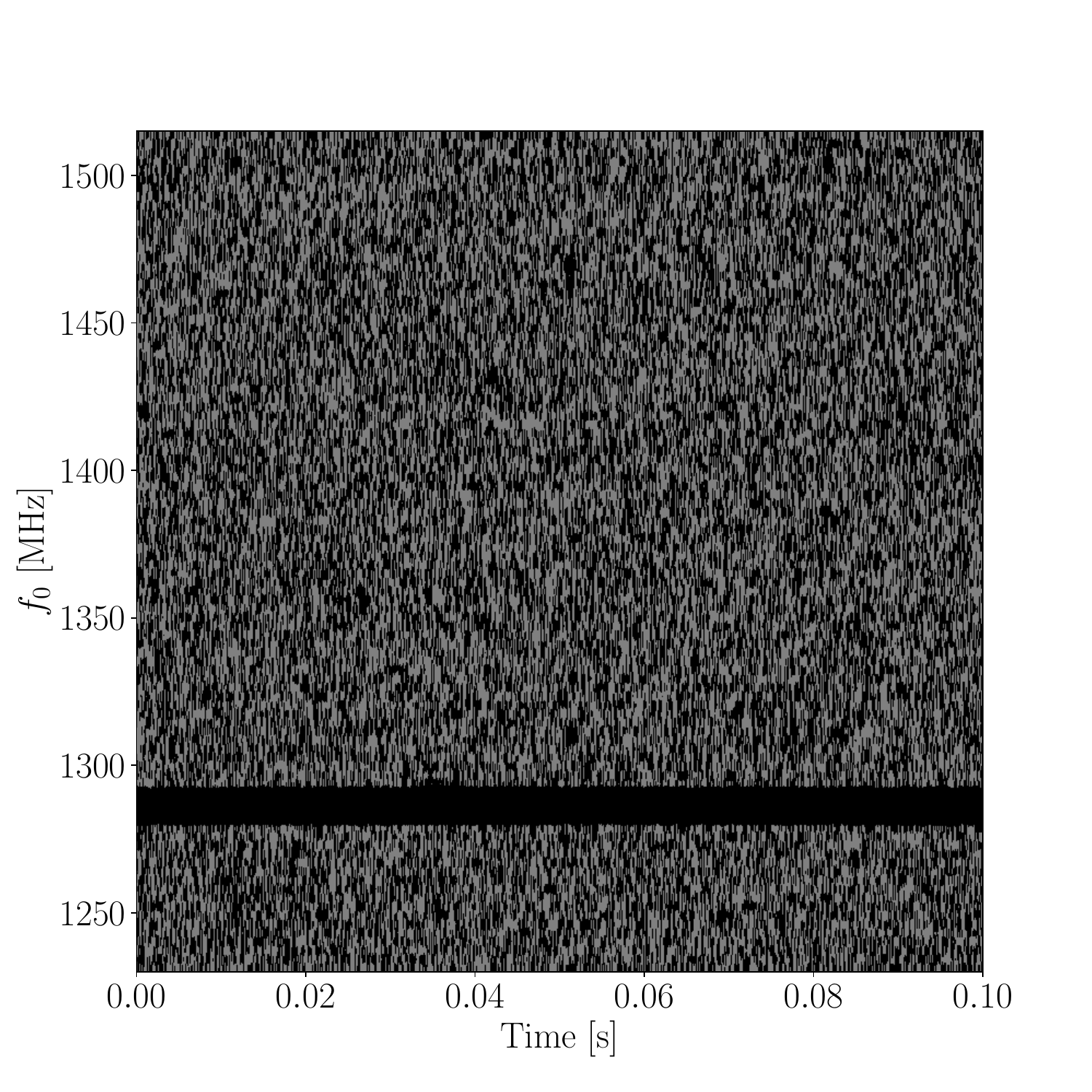}}
    \caption{Same as Figure \ref{Fig:specNoRFI} but with narrowband, impulsive RFI injected using the \texttt{simulateRFI} subroutine of the \texttt{simulatesearch} software package \citep{Luo_2022}. The injected narrowband RFI is visible as a black band between 1280 MHz and 1290 MHz.}
    \label{Fig:specRFI}
\end{figure}

The validation and performance tests in Sections \ref{Sec:Validation} and \ref{Sec:PerformanceTests} assume that RFI is excised from the synthetic radio survey data generated in this paper. In real radio survey data, however, the effects of RFI are mitigated across several complex data processing stages \citep{Knispel_2013,Ng_2015,VanHeerdan_2017,Sobey_2022}. In this appendix, we assess the HMM response to narrowband, impulsive RFI. That is, we inject RFI into the Gaussian synthetic survey data in Section \ref{SubSec:SyntheticData} using the \texttt{simulateRFI} subroutine of the \texttt{simulatesearch} software package \citep{Luo_2022}. The injected RFI is visible in the spectrogram in Figure \ref{Fig:specRFI} as a black band between 1280 MHz and 1290 MHz. We then repeat the analysis in Section \ref{SubSec:FrequencyTracking}. To compare the HMM response to the synthetic RFI-free data in Figure \ref{Fig:specNoRFI} with the RFI-affected data of Figure \ref{Fig:specRFI} fairly, we fix the random seed employed by the \texttt{simulateSystemNoise} subroutine of the \texttt{simulatesearch} software package.  We refer the reader to Section 3.2 of \cite{Luo_2022} for details about the different types of RFI that can be simulated using the \texttt{simulatesearch} software package. 

\begin{figure}
\centering{
    \includegraphics[width=0.8\textwidth, keepaspectratio]{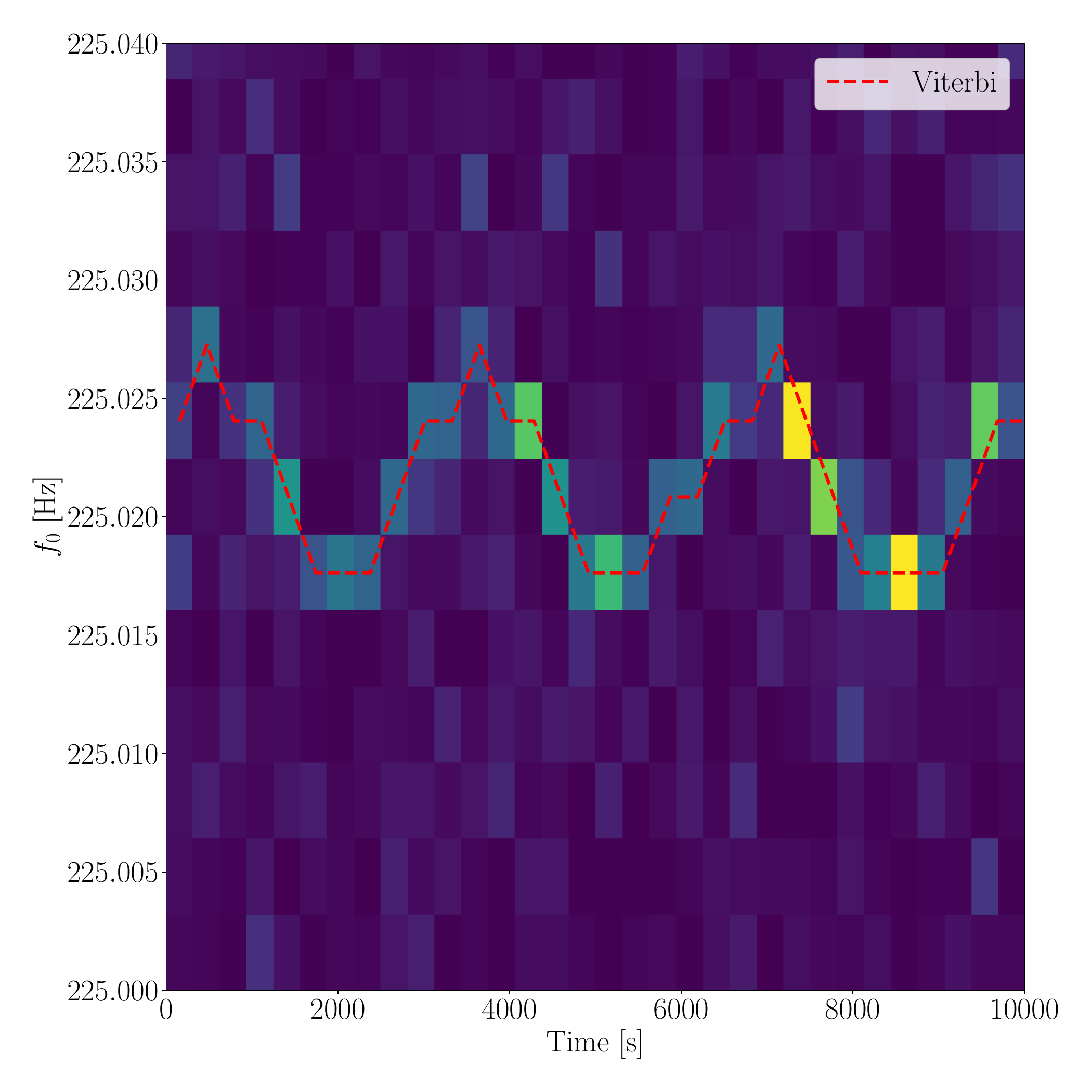}}
    \caption{Frequency tracking results for the representative test source in Section \ref{Sec:Validation} in a single subband $220 \leq f_0 / (1\, {\rm Hz}) \leq 230$ which contains an above-threshold outlier with ${\cal L} > {\cal L}_{\rm th}$ and narrowband RFI, generated synthetically using the \texttt{simulatesearch} software package. Magnified subset of the frequency-time spectrogram with 32$\times$12 pixels, whose coloring indicates as a heat map the value of the normalized Fourier power, calculated according to Equation (\ref{Eq:Power}). The red, dashed curve is the optimal hidden state sequence $f_{\rm p}(t)$ output by the Viterbi algorithm.}
    \label{Fig:RFIPlots}
\end{figure}

In Figure \ref{Fig:RFIPlots} we present the Viterbi frequency tracking results for the $220 \leq f_0/(1 \, {\rm Hz}) \leq 230$ subband in the RFI-affected data. In the top panel, we plot $\mathcal{L} = \ln P(Q^*|O)$ versus the observed frequency $f_0$ as a black curve. Overplotted as a gray, horizontal, dashed line is the likelihood threshold $\mathcal{L}_{\rm th}$ calculated in Section \ref{SubSec:NoiseOnly}. The blue, vertical, dashed line corresponds to the injected pulse frequency $f_{\rm p, inj}$ in Table \ref{tab:SourceParameters}. In the bottom panel, we plot the resulting $N_T \times N_Q = 32 \times 3120$ bins of the time-frequency spectrogram. For each bin in the time-frequency plane, the coloring indicates the value of the detection statistic, i.e. the normalized Fourier power, calculated according to Equation (\ref{Eq:Power}), with brighter colors indicating a higher value in the same fashion as Figure \ref{Fig:specNoRFI}. Overplotted as a dashed, red curve is the optimal hidden state sequence $f_{\rm p }(t)$, constructed according to Steps \ref{Alg:Backtrack1}--\ref{Alg:Backtrack3} in Algorithm \ref{Alg:Viterbi}.

The log-likelihood of the optimal Viterbi path is $\mathcal{L} = \ln P(Q^*|O) = 334.79$, to be compared with $\mathcal{L} = 345.18$ in Section \ref{SubSec:FrequencyTracking} in the absence of RFI. That is, the log-likelihood of the optimal Viterbi path decreases in sensitivity by $13.39$ when the sequence of observations $O$ ingested by the HMM contain impulsive narrowband RFI. The results are encouraging. For example, the detection is significant; the probability of $\mathcal{L} = \ln P(Q^*| O) = 334.79$ occurring by chance is less than $10^{-20}$. We remind the reader that the results in this appendix are presented as a rudimentary starting point only. A fuller study of the HMM response to RFI is postponed to future work. 

\section{Likelihood threshold}\label{App:sigma}

The likelihood threshold $\mathcal{L}_{\rm th}$ introduced in Section \ref{SubSec:Threshold} and set in Section \ref{SubSec:NoiseOnly} is a complicated, nonlinear function of the random variables $\Tilde{\lambda}$ and $N_{\rm tail}$, i.e. $\mathcal{L}_{\rm th} = \mathcal{L}_{\rm th}(\Tilde{\lambda}, N_{\rm tail})$. Accordingly, it is challenging to write down explicitly important statistical quantities such as ${\rm{Var}}(\mathcal{L}_{{\rm th}}) = \langle (\delta \mathcal{L}_{\rm th})^2 \rangle$, with $\delta \mathcal{L}_{\rm th} = \mathcal{L}_{\rm th} - \langle \mathcal{L}_{\rm th} \rangle$, because the sampling distribution $p(\mathcal{L}_{\rm th})$ is not known in closed form. In this appendix, we adopt the first-order delta method \citep{Casella_2024} to approximate $\langle (\delta \mathcal{L}_{\rm th})^2 \rangle$ and verify the approximation using statistical bootstrapping \citep{Mooney_1993}. In Appendix \ref{SubSec:FirstOrderDelta} we introduce the  first-order delta method as an approximate approach to estimate the variance of  $g(\bm{\Theta})$, a smooth, differentiable, nonlinear function of $\bm{\Theta}$, where $\bm{\Theta} = (\Theta_1, \hdots , \Theta_M)$ denotes an arbitrary $M$-dimensional vector of random variables. In Appendix \ref{SubSec:SigmaLDer} we specialize the first-order delta method, introduced in Appendix \ref{SubSec:FirstOrderDelta}, to estimate $\sigma_{\mathcal{L}_{\rm th}}^2 = \langle (\delta \mathcal{L}_{\rm th})^2 \rangle$. In Appendix \ref{SubSec:BootStrapping} we verify  $\sigma_{\mathcal{L}_{\rm th}}^2 = \langle (\delta \mathcal{L}_{\rm th})^2 \rangle$ empirically using bootstrap resampling \citep{Mooney_1993}.

\subsection{First-order delta method}\label{SubSec:FirstOrderDelta}

The first-order delta method is a powerful technique adopted in statistics to approximate the mean $\langle g \rangle$ and variance $\langle (\delta g)^2 \rangle$ of the random variable $g = g(\bm{\Theta})$.  It proceeds as follows. Let the (finite) mean and variance of $\bm{\Theta} = (\Theta_1,\hdots,\Theta_{M})$ exist and be defined according to 
\begin{equation}
    \bm{\mu} = \langle \bm{\Theta} \rangle,
\end{equation}
and 
\begin{equation}\label{Eq:GenSigma}
    \bm{\Sigma} = \langle (\bm{\Theta} - \bm{\mu})(\bm{\Theta} - \bm{\mu})^{\rm T} \rangle,
\end{equation}
respectively. In Equation (\ref{Eq:GenSigma}), the superscript `T' denotes a matrix transpose. To first order, one has 
\begin{equation}
    g(\bm{\Theta}) = g(\bm{\mu}) + \nabla g(\bm{\mu})^{\rm T} \left(\bm{\Theta} - \bm{\mu} \right),
\end{equation}
where $\nabla \equiv (\partial/\partial \Theta_1, \hdots, \partial/\partial \Theta_{M})$. The mean $\langle g \rangle$ and variance $\langle (\delta g)^2 \rangle$ are given respectively by
\begin{equation}
    \langle g \rangle = g(\bm{\mu}),
\end{equation}
and  
\begin{equation} \label{Eq:VarianceFOD}
    \langle (\delta g)^2 \rangle = \nabla g(\bm{\mu})^{\rm T} \, \bm{\Sigma} \, \nabla g(\bm{\mu}).
\end{equation}
We refer the reader to Section 5.5.4 of \cite{Casella_2024} for further details about deriving Equation (\ref{Eq:VarianceFOD}).

\subsection{Deriving $\mathcal{\sigma_{\mathcal{L}_{\rm th}}}$}\label{SubSec:SigmaLDer}

In the context of discovering pulsars in compact binaries, the uncertainty associated with the likelihood threshold $\mathcal{L}_{\rm th}$ in Sections \ref{SubSec:Threshold} and \ref{SubSec:NoiseOnly} is approximated to first order using Equation (\ref{Eq:VarianceFOD}). Specifically, one has 
\begin{equation}\label{Eq:DeltaLth}
    \nabla \mathcal{L}_{\rm th}(\Tilde{\lambda}, N_{\rm tail}) = [\Tilde{\lambda}^{-2} \log\left(C/N_{\rm{tail}}\right), ( \Tilde{\lambda} N_{\rm{tail}})^{-1}]^{\rm{T}},
\end{equation}
and 
\begin{equation}\label{Eq:CovMatrix}
    \bm{\Sigma} =   \begin{bmatrix}
                        \langle(\delta   \Tilde{\lambda})^2\rangle & 0 \\
                        0 & \langle\left(\delta N_{\rm tail} \right)^2\rangle
                    \end{bmatrix},
\end{equation}
with $C = N_{\rm{real}} N_{Q}  [1 - (1 - \alpha')^{1/N_{Q}}]$. We draw the reader's attention to two important simplifying assumptions about Equations (\ref{Eq:DeltaLth}) and (\ref{Eq:CovMatrix}). First, we evaluate $\nabla \mathcal{L}_{\rm th}(\Tilde{\lambda}, N_{\rm tail})$ using the maximum likelihood estimate $\Tilde{\lambda}$ [see Equation (\ref{Eq:MLELambda})] and $N_{\rm tail}$, instead of $\langle \lambda \rangle$ and $\langle N_{\rm tail} \rangle$, because the latter are unknown in practice. Second, we assume that $\Tilde{\lambda}$ and $N_{\rm tail}$ are independent, i.e.\ the covariance terms in Equation (\ref{Eq:CovMatrix})  such as $\langle (\delta \Tilde{\lambda}) \, (\delta N_{\rm tail}) \rangle = 0$, for $\delta \Tilde{\lambda} = \Tilde{\lambda} - \langle \Tilde{\lambda} \rangle$ and $\delta N_{\rm tail} = N_{\rm tail} - \langle N_{\rm tail} \rangle$. The latter assumption is justified empirically in Appendix \ref{SubSec:BootStrapping} below and can be relaxed in the future if necessary.

The variances in Equation (\ref{Eq:CovMatrix}), $\langle (\delta \Tilde{\lambda})^2\rangle$ and $\langle (\delta N_{\rm tail})^2 \rangle$, are estimated as follows. We adopt a large-sample, asymptotically Gaussian approach to the maximum likelihood estimate $\Tilde{\lambda}$ using the Fisher information $\mathcal{I}_{N_{\rm tail}}(\Tilde{\lambda}) = N_{\rm tail}/\Tilde{\lambda}^2$ for $N_{\rm tail}$ independent, identically distributed samples, i.e.\  $\Tilde{\lambda} \sim \mathcal{N}( \Tilde{\lambda},1/\mathcal{I}_{N_{\rm tail}}) \sim \mathcal{N}(\Tilde{\lambda}, \Tilde{\lambda}^2/N_{\rm tail})$, and hence $\langle (\delta \Tilde{\lambda})^2 \rangle \approx \Tilde{\lambda}^2/N_{\rm tail}$ \citep{Newey_1994,VanDerVaart_2000,Casella_2024}. Similarly we treat the tail count $N_{\rm tail}$ as a Poisson random variable, i.e.\ a large-sample, rare-event Binomial random variable, with $\langle (\delta N_{\rm tail})^2 \rangle \approx N_{\rm tail}$; see Section 2.2 of \cite{Cowan_1998} for further details. Accordingly, the variance of the likelihood threshold $\langle (\delta \mathcal{L}_{\rm th})^2 \rangle$ is given by 
\begin{equation}\label{Eq:LthVariance}
     \langle (\delta \mathcal{L}_{\rm th})^2 \rangle = \left[ 1 +  \log^2 \left( C / N_{\rm tail} \right) \right]/( \Tilde{\lambda}^2 N_{\rm tail}).
\end{equation} 
Hence, the first-order uncertainty of the detection threshold $\mathcal{L}_{\rm th}$ set in Section \ref{SubSec:NoiseOnly} is given by $\sigma_{\mathcal{L}_{\rm th}} = \langle (\delta \mathcal{L}_{\rm th})^2 \rangle^{1/2} = 0.052$.

\subsection{Bootstrap resampling}\label{SubSec:BootStrapping}

Bootstrap resampling is a computationally efficient, nonparametric technique to approximate $\sigma_{\mathcal{L}_{\rm th}} = \langle (\delta \mathcal{L}_{\rm th})^2 \rangle^{1/2}$. It treats the empirical distribution of the exponential tail likelihood values as an approximation to the unknown true noise distribution, and propagates finite-sample fluctuations through the full threshold-estimation pipeline described in Section \ref{SubSec:Threshold}. Specifically, we generate $\mathcal{B}$ bootstrap realizations by sampling with replacement from $\{X_i = \mathcal{L}_i - \mathcal{L}_{\rm tail}\}_{i = 1}^{N_{\rm tail}}$. For each bootstrap realization $b$ we recompute $\Tilde{\lambda}^{(b)}$ and hence $\mathcal{L}^{(b)}_{\rm th}$ using the same false-alarm prescription as Section \ref{SubSec:NoiseOnly}.

\begin{figure}
\centering{
    \includegraphics[width=0.8\textwidth, keepaspectratio]{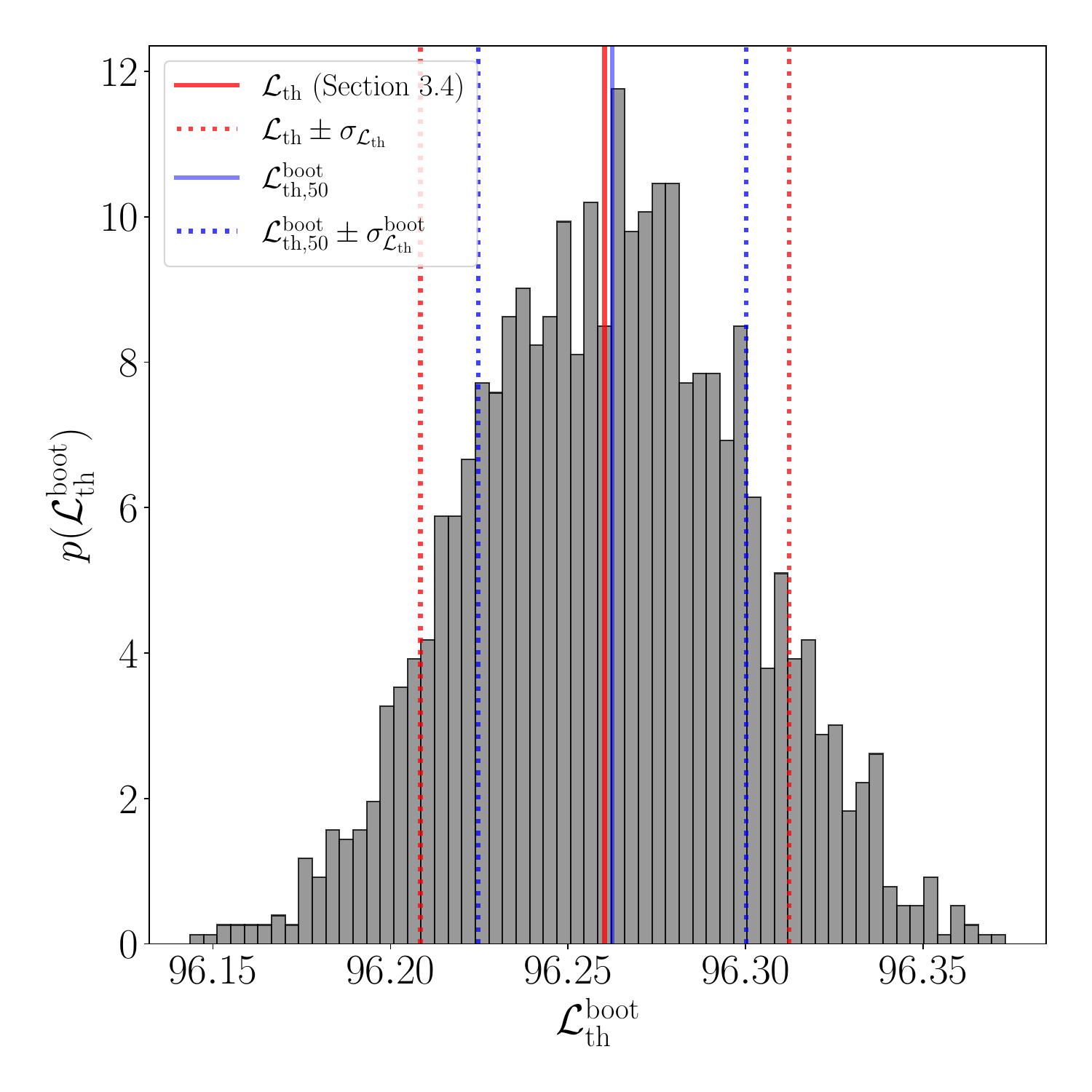}}
    \caption{Probability density $p(\mathcal{L}_{\rm th}^{\rm boot})$ (gray histogram) of $B=2000$ bootstrap realizations of $\mathcal{L}_{\rm th}^{\rm boot}$. Overplotted as red and blue lines are the likelihood threshold $\mathcal{L}_{\rm th}$ set in Section \ref{SubSec:NoiseOnly} and $\mathcal{L}_{{\rm th},50}^{\rm boot}$, the median value of $p(\mathcal{L}_{\rm th}^{\rm boot})$, respectively. The uncertainties associated with $\mathcal{L}_{\rm th}$ and $\mathcal{L}_{{\rm th},50}^{\rm boot}$ are overplotted as $\mathcal{L}_{\rm th} \pm \sigma_{\mathcal{L}_{\rm th}}$ and $\mathcal{L}_{{\rm th},50}^{\rm boot} \pm \sigma_{\mathcal{L}_{\rm th}}^{\rm boot}$ bracketed by dashed, red and blue lines, respectively.}
    \label{Fig:BootstrapDistribution}
\end{figure}

In Figure \ref{Fig:BootstrapDistribution} we plot the bootstrapped likelihood threshold distribution $p(\mathcal{L}_{\rm th}^{\rm boot})$ as a gray histogram using $\mathcal{B} = 2000$ bootstrap realizations of $\mathcal{L}_{\rm th}^{\rm boot}$. We plot the likelihood threshold $\mathcal{L}_{\rm th}$ set in Section \ref{SubSec:NoiseOnly} and the median value of the bootstrapped likelihood threshold distribution $\mathcal{L}_{{\rm th},50}^{\rm boot}$ using red and blue lines, respectively. The uncertainties associated with $\mathcal{L}_{\rm th}$ and $\mathcal{L}_{{\rm th},50}^{\rm boot}$ are overplotted as $\mathcal{L}_{\rm th} \pm \sigma_{\mathcal{L}_{\rm th}}$ and $\mathcal{L}_{{\rm th},50}^{\rm boot} \pm \sigma_{\mathcal{L}_{\rm th}}^{\rm boot}$ bracketed by dashed, red and blue lines, respectively. The analysis reveals that the uncertainty approximated using bootstrap resampling, i.e.\ $\sigma_{\mathcal{L}_{\rm th}}^{\rm boot} = 0.037$, is in broad agreement with $\sigma_{\mathcal{L}_{\rm th}} = \langle (\delta \mathcal{L}_{\rm th})^2 \rangle^{1/2} = 0.052$ of Section \ref{SubSec:FirstOrderDelta}.

\section{Secular pulsar braking}\label{App:PulsarBrak}
The tests in Sections \ref{Sec:Validation} and \ref{Sec:PerformanceTests} assume that the injected pulse frequency is constant apart from Doppler modulation, e.g.\ $f_{\rm p,inj} = 225.02 \, \rm{Hz}$ in the pulsar rest frame, as in Table \ref{tab:SourceParameters}. It is natural to ask whether the secular spin-down rate of the pulsar, i.e.\ $\dot{f}_{\rm p, inj}$, affects the performance of the HMM. Theoretically one expects not: the transition probability specified in Equation (\ref{Eq:TransSpec}) accommodates a decrement in the hidden state $f_{\rm p}(t)$ from one time step to the next irrespective of the cause (secular spin down or Doppler modulation). The total change in rotational phase caused by Doppler modulation exceeds that caused by $\dot{f}_{\rm p,inj}\neq 0$, because $\dot{f}_{\rm p, inj}$ always satisfies Equation (\ref{Eq:frequencyDer}) for the pulsars targeted in this paper.

To verify the above claim empirically, we repeat the analysis in Section \ref{Sec:Validation} with $\dot{f}_{\rm p,inj} = -9.8\times 10^{-13} \, \rm{s^{-2}}$, the secular spin-down rate measured for PSR J1906$+$0746 \citep{Lorimer_2006}. Specifically, we follow the same procedure as the validation tests in Section \ref{Sec:Validation} with two minor exceptions. First, we reduce the number of simulated frequency channels in the synthetic pulsar survey data in Section \ref{SubSec:SyntheticData} from 96 to eight, to reduce the computational overhead. Second, we restrict attention to the optimal Viterbi path $\mathcal{L} = \ln P(Q^*|O)$ in the $220 \leq f_0/(1 \, \rm{Hz}) \leq 230$ subband. We find (results not plotted for brevity) that the log-likelihood of the optimal Viterbi path, $\mathcal{L} = \ln P(Q^*|O) = 371.27 > \mathcal{L}_{\rm th} = 96.2 \pm 0.052$, peaks within two bins ($0.0064 \, \rm{Hz}$) of the injected pulse frequency $f_{\rm p, inj} = 225.02 \, \rm{Hz}$ for $\dot{f}_{\rm p,inj} = -9.8\times 10^{-13} \, \rm{s^{-2}}$, as expected. 

\label{lastpage}
\newpage
\bibliography{main}{}
\bibliographystyle{aasjournalv7}

\end{document}